\documentclass[]{iopart}

\usepackage{graphicx}
\usepackage{bm}
\usepackage[colorlinks=true,linkcolor=blue,citecolor=blue]{hyperref}

\begin{document}

\title[Semi-classical {\em vs.} quantum description of the ground state of three-level atoms\dots ]{Semi-classical {\em vs.} quantum description of the ground state of three-level atoms interacting with a one-mode electromagnetic field}

\author{S Cordero, O Casta\~nos,  R L\'opez-Pe\~na and E~Nahmad-Achar} 

\address{%
Instituto de Ciencias Nucleares, Universidad Nacional Aut\'onoma de M\'exico, Apartado Postal 70-543, 04510 M\'exico DF,   Mexico  \\ }

\ead{\mailto{sergio.cordero@nucleares.unam.mx},\,\mailto{ocasta@nucleares.unam.mx},\\ \mailto{lopez@nucleares.unam.mx},\,\mailto{nahmad@nucleares.unam.mx}}

\date{\today}

\begin{abstract}
We consider $N_a$ three-level atoms (or systems) interacting with a one-mode electromagnetic field in the dipolar and rotating wave approximations.  The order of the quantum phase transitions is determined explicitly for each of the configurations $\Xi$, $\Lambda$ and $V$, with and without detuning.  The semi-classical and exact quantum calculations for both the expectation values of the total number of excitations $\cal{M}=\langle \bm{M} \rangle$ and photon number $n=\langle \bm{n} \rangle$ have an excellent correspondence as functions of the control parameters. We prove that the ground state of the collective regime obeys sub-Poissonian statistics for the ${\cal M}$ and $n$ distribution functions. Therefore, their corresponding fluctuations are not well described by the semiclassical approximation. We show that this can be corrected by projecting the variational state to a definite value of ${\cal M}$.
\end{abstract}

\pacs{42.50.Ct,42.50.Nn,73.43.Nq,03.65.Fd}%

\maketitle

\section{Introduction}

Interaction of $N_a$ two-level atoms with a quantized electromagnetic field, using dipolar and rotating wave approximations, is described by the {\em Tavis-Cummings Model} \cite{tavis68,tavis69}, having an extensive use in quantum optics \cite{dodonov03}. Recently this model has been
physically realized using a QED cavity with Bose-Einstein
condensates \cite{baumann10,nagy10}. Particularly interesting has
been the investigation of the phase transitions of the
system in the thermodynamic limit \cite{hepp73,wang73}, and at zero
temperature \cite{buzek05,castanos09a,castanos09b}.

The system of three-level atoms interacting with a one mode radiation field together with a dipole-dipole interaction between the atoms has been studied to determine the atomic squeezing~\cite{civitarese1,civitarese2}. They consider $\Xi$ and $\Lambda$ configurations under initial conditions of the matter and field parts associated to $SU(2)$ and Heisenberg-Weyl coherent states, respectively.  Spin variances for the $V$ and $\Lambda$ configurations of an ensemble of atoms interacting with two light fields, a coherent pump state and a squeezed vacuum as a probe, have been calculated by means of the Langevin equations derived from the Bloch equations~\cite{dantan}. By using a Holstein-Primakoff mapping, two stable states, normal and superradiant (the latter in two colors), have been identified in the thermodynamic limit for the $\Lambda$ configuration~\cite{brandes}. 

More recently, we have analytically obtained the localization of the quantum phase transitions from the normal to the collective regimes for three-level atoms interacting with a one-mode field for the $\Xi$, $\Lambda$ , and $V$ configurations, in the rotating wave approximation (RWA). These transitions appear in the ground state energy surface ${\cal E}^c$ and the corresponding total number of excitations ${\cal M}^c$, when plotted as functions of their corresponding dipole coupling constants (control parameters), calculated using as test function the direct product of the Heisenberg-Weyl (field contribution) and  Gelfand-Tsetlin (matter contribution) coherent states. We found that the agreement of these quantities with the corresponding exact quantum calculations (namely $E^q$ and $M^q$) is remarkable~\cite{clpcna13}. 

In this paper we determine explicitly the order of the quantum phase transitions, and calculate the Mandel parameter of the ${\cal M}$ distribution function and of the photon number distribution function of the ground state of the system. We find that first- and second-order transitions appear for atoms in the $\Xi$ configuration, and only second-order transitions appear for atoms in the $V$ configuration. Atoms in the $\Lambda$ configuration, depending of the detuning parameter, mimic the behavior of the $\Xi$ or the $V$ configuration. We find that in the collective regime, i.e., where the ground state possesses ${\cal M} >0$, the state obeys sub-Poissonian statistics while in the normal regime it satisfies Poissonian statistics. 

While both, the total number of excitations ${\cal M}$ and the expectation value of the number of photons $\langle \bm{n}\rangle$, are in agreement with their corresponding exact quantum calculation, we find that their fluctuations are not. This is because the semi-classical ground state has the contribution of an infinite number of photons in a Poissonian distribution. The above suggest a projection of the test function to a definite value of the total number of excitations. This we do by means of a discretization of ${\cal M}$, according to its expectation value with respect to the test function. We prove that this {\it projected state} provides the appropriate correction, where now ${\cal M}$, $\langle \bm{n}\rangle$, and their corresponding fluctuations are in excellent agreement with the exact quantum calculation.

The paper is organized as follows: Sec.~\ref{formalism} presents in general the problem for $N_a$ atoms of $N$-levels interacting with $L$-modes of a quantized electromagnetic field in the dipolar approximation.  In Sec.~\ref{three.level.atoms} we restrict to the problem of three-level atoms interacting with a one-mode quantized electromagnetic field (QEMF) in the RWA approximation, and establish the corresponding constant of motion ${\cal M}$ (total number of excitations) for each atomic configuration. In \ref{GCS} the test function as a direct product of Heisenberg-Weyl (field contribution) and Gelfand-Tsetlin (matter contribution) coherent states is proposed for the semi-classical approximation. The corresponding semi-classical energy of the problem is calculated in \ref{energy_surface}. In \ref{order.trans} we provide an exact expression to evaluate the first order derivatives of the ground state energy surface (as a function of the control parameters), so that the first-order  transitions for each atomic configuration can be calculated in analytical form.  For every value of the total number of excitations, the corresponding Mandel parameter of the semi-classical ground state, providing the kind of statistics that it satisfies, is evaluated in \ref{fluc}.  In \ref{numerical.results} we show the numerical results for both order transitions, the Mandel parameter and the photon expectation value, for all different atomic configurations.  Sec.~\ref{QNS} presents the exact quantum calculations and compares them with the semi-classical ones. In Sec. \ref{quantum.proj} the calculations obtained by using the projected variational state with the corresponding exact quantum results are compared. Finally, we give in Section \ref{concluding} some concluding remarks.

\section{$N$-level atoms interacting with an $L$-mode QEMF}\label{formalism}

We consider, in the dipolar approximation, the Hamiltonian of $N_a$ identical atoms of $N$-levels interacting with $L$-modes of a quantized electromagnetic field. Let $\bm{A}_{ij}^{(k)}$ denote the atomic operator of the $k$-th atom. For each atom, these operators obey a unitary algebra ${\rm u}_k(N)$ in $N$ dimensions, i.e.,
%
%\begin{subequations}
\begin{eqnarray}
\sum_{i=1}^{N} \bm{A}_{ii}^{(k)} = 1,\\
\left[\bm{A}_{ij}^{(k)},\bm{A}_{lm}^{(k')}\right] = \delta_{kk'}\left(\delta_{jl}\bm{A}_{im}^{(k)} - \delta_{im}\bm{A}_{lj}^{(k)}\right).
\end{eqnarray}
%\end{subequations}
%
Defining 
\begin{eqnarray}
\bm{A}_{ij} &\equiv& \sum_{k=1}^{N_a} \bm{A}_{ij}^{(k)},\label{op.Aij} 
\end{eqnarray}
one can see that the following relationships are fulfilled 
%
%\begin{subequations}
\begin{eqnarray}
\bm{n}_a= \sum_{i=1}^{N} \bm{A}_{ii}  \label{eq.Na}\\
\left[\bm{A}_{ij},\bm{A}_{lm}\right] = \delta_{jl}\bm{A}_{im} - \delta_{im}\bm{A}_{lj}. \label{eq.AijAlm}
\end{eqnarray}
%\end{subequations}
% 
We have here defined the operator $\bm{n}_a$ representing the total number of atoms, with eigenvalue $N_a$, and Eq. (\ref{eq.AijAlm}) shows that the set of operators $\bm{A}_{ij}$ obey the commutation relations of a unitary algebra in $N$ dimensions,  ${\rm u}(N)=  \oplus^{N_a}_{k=1} \, u_k(N)$.

Now, for $L$-modes of a quantized field and $N_a$ atoms, the free Hamiltonian may be written as ($\hbar = 1$)
\begin{eqnarray}
\bm{H}_0 = \sum_{\ell = 1}^L \Omega_\ell \bm{a}_\ell^\dag \bm{a}_\ell + \sum_{i=i}^N\omega_i\bm{A}_{ii},
\end{eqnarray}
where $\Omega_\ell$ and $\omega_i$ correspond, respectively, to the frequencies of the $\ell$-th field mode and $i$-th atomic level (we choose $\omega_1\leq\omega_2\leq\cdots\leq\omega_N$). Here $\bm{a}_\ell^\dag,\ \bm{a}_\ell$ are the usual creation and annihilation operators of the field obeying the boson algebra, i.e.,
\begin{eqnarray}
\left[\bm{a}_i,\bm{a}_j^\dag \right] = \delta_{ij},  
\end{eqnarray}
and  $\bm{A}_{ij}$ are the atomic operators of Eq. (\ref{op.Aij}).

The interaction Hamiltonian due to the dipole operator $\vec{\bm{d}}$ of the atoms with the electromagnetic field $\vec{\bm{E}}$, reads as \cite{haroche}
\begin{eqnarray}\label{eq.DE}
\bm{H}_{int} = -\vec{\bm{d}}\cdot\vec{\bm{E}}.
\end{eqnarray}
$\vec{\bm{d}}$ may be written as
\begin{eqnarray}\label{eq.D}
\vec{\bm{d}} = \sum_{i\neq j} \vec{d}_{ij} \bm{A}_{ij}, 
\end{eqnarray} 
where $\vec{d}_{ij}$ represent the matrix elements of the vector dipole operator between the levels $j$ and $i$. Notice that $\vec{\bm{d}}$ has no diagonal contributions, because the dipolar interaction of a level with itself is zero.   The corresponding quantized field may be written as
\begin{eqnarray}\label{eq.E}
\vec{\bm{E}} = \sum_{\ell=1}^L \left[\vec{{\cal E}}_\ell(\vec{r})\bm{a}_\ell + \vec{{\cal E}}^*_\ell(\vec{r})\bm{a}_\ell^\dag\right],
\end{eqnarray}
where $\vec{{\cal E}}_\ell(\vec{r})$ obeys the Helmholtz equation for the $\ell$-th field mode, providing the structure of the field into the cavity. Substituting Eqs. (\ref{eq.D}) and (\ref{eq.E}) into Eq. (\ref{eq.DE}), and reordering the different contributions, one may write the interaction Hamiltonian as
\begin{eqnarray}\label{eq.Hint.1.full}
\bm{H}_{int} &=& -\sum_{s=1}^{N-1}\sum_{\ell=1}^L \left(\bm{a}_\ell^\dag {\vec{g}_{s\ell}}\cdot\vec{\bm{\sigma}}_{s-} + \bm{a}_\ell \vec{\bm{\sigma}}_{s+}\cdot {\vec{g}_{s\ell}}^{\, *\textrm{\tiny T}} \right)  \nonumber \\
&-& \sum_{s=1}^{N-1}\sum_{\ell=1}^L \left(\bm{a}_\ell {\vec{g}_{s\ell}}^{\, *}\cdot\vec{\bm{\sigma}}_{s-} + \bm{a}_\ell^\dag \vec{\bm{\sigma}}_{s+}\cdot {\vec{g}_{s\ell}}^{\, \textrm{\tiny T}} \right),\qquad
\end{eqnarray}
where were defined the vector operators
\begin{eqnarray}
\vec{\bm{\sigma}}_{s+} = \left(\bm{A}_{1+s,1},\dots,\bm{A}_{j+s,j},\dots, \bm{A}_{(N-s)+s,N-s}\right)\quad
\label{sigmaop}
\end{eqnarray}
containing the set of operators $\bm{A}_{ij}$ with transitions from the $j$-th level of the atom to the $(j+s)$-th level. Also, $\vec{\bm{\sigma}}_{s-} = \vec{\bm{\sigma}}_{s+}^\dag$, and
\begin{eqnarray}
\vec{g}_{s\ell} = \frac{1}{\sqrt{N_a}}\left(\mu_{1,1+s}^{(\ell)}, \dots, \mu_{j,j+s}^{(\ell)}, \dots, \mu_{N-s,(N-s)+s}^{(\ell)}\right)\qquad \label{eq.gsk}
\end{eqnarray}    
with $\mu_{ij}^{(\ell)}/\sqrt{N_a} = \vec{d}_{ij}\cdot \vec{{\cal E}}_\ell^*$, the coupling parameter between levels $i$ and $j$, and where we have taken $\vec{d}_{ji}=\vec{d}_{ij}$. Here, we have eliminated the dependence on $\vec{r}$ of  $\vec{{\cal E}}_k^*$ by supposing that the $N_a$ atoms are stationary at the center of the cavity, and that the field is a smooth function in that region.  

The second term in the rhs of equation (\ref{eq.Hint.1.full}) corresponds to the counter-rotating term, and when RWA approximation is considered this term is neglected. So the interaction term in the RWA approximation is given by
\begin{eqnarray}\label{eq.Hint.1}
\bm{H}_{int} = -\sum_{s=1}^{N-1}\sum_{\ell=1}^L \left(\bm{a}_\ell^\dag {\vec{g}_{s\ell}}\cdot\vec{\bm{\sigma}}_{s-} + \bm{a}_\ell \vec{\bm{\sigma}}_{s+}\cdot {\vec{g}_{s\ell}}^{\, *\textrm{\tiny T}} \right). \quad
\end{eqnarray}
Finally, the full Hamiltonian in RWA reads as
\begin{eqnarray}\label{eq.H.full}
\bm{H} &=& \sum_{\ell = 1}^L \Omega_\ell \bm{a}_\ell^\dag \bm{a}_\ell + \sum_{j=1}^N\omega_j\bm{A}_{jj}  \nonumber \\ 
&-& 
\sum_{s=1}^{N-1}\sum_{\ell=1}^L \left(\bm{a}_\ell^\dag {\vec{g}_{s\ell}}\cdot\vec{\bm{\sigma}}_{s-} + \bm{a}_\ell \vec{\bm{\sigma}}_{s+}\cdot {\vec{g}_{s\ell}}^{\, *\textrm{\tiny T}} \right).
\end{eqnarray}
The Hamiltonian above shows the underlying structure of the unitary group in $N$ dimensions, $U(N)$, which makes natural the use of the Gelfand-Tsetlin states~\cite{gelfand}. This allows for the description, in general, of systems with any kind of symmetry, including distinguishable particles.

\section{Three-level atoms interacting with a one-mode QEMF}\label{three.level.atoms}

In what follows we consider $N_a$ three-level atoms interacting with a one-mode QEM field, i.e., we choose $N=3$ and $L=1$ in Eq. (\ref{eq.H.full}).  Replacing the corresponding values of $\vec{\bm{\sigma}}_{s\pm}$ and $\vec{g}_{s\ell}$ into Eq. (\ref{eq.H.full}) one finds the Hamiltonian of the system as
\begin{eqnarray}\label{eq.H.3level}
\bm{H} &=& \Omega \bm{a}^\dag\bm{a} + \omega_1\bm{A}_{11} + \omega_2\bm{A}_{22} + \omega_3\bm{A}_{33}  %\nonumber \\
- \frac{1}{\sqrt{N_a}}\mu_{12}\left(\bm{a}\bm{A}_{21} + \bm{a}^\dag\bm{A}_{12} \right)\nonumber \\ 
&-&\frac{1}{\sqrt{N_a}}  \mu_{13}\left(\bm{a}\bm{A}_{31} + \bm{a}^\dag\bm{A}_{13}\right) %\nonumber \\
-\frac{1}{\sqrt{N_a}} \mu_{23}\left(\bm{a}\bm{A}_{32} + \bm{a}^\dag\bm{A}_{23}\right) \, .
\end{eqnarray}
where the subscript on the field operators is no longer necessary, and without loss of generality we assume that the coupling constants obey $\mu_{ij} = \mu_{ij}^* = \mu_{ji}$. The only operator of the form  $\bm{C} = \lambda\bm{a}^\dag\bm{a} + \lambda_1\bm{A}_{11} + \lambda_2\bm{A}_{22} + \lambda_3\bm{A}_{33}$ that commutes with the Hamiltonian Eq. (\ref{eq.H.3level}) is given by Eq. (\ref{eq.Na}), i.e., the total number of atoms is conserved. However, if one allows one coupling term $\mu_{ij}$ to be zero, it is possible to find another operator that commutes with the Hamiltonian Eq. (\ref{eq.H.3level}). This operator, for each atomic configuration, is given by
%
%\begin{subequations}\label{eq.M}
\begin{eqnarray}
\bm{M}_{\Xi} = \bm{a}^\dag\bm{a} + \bm{A}_{22} + 2\bm{A}_{33} \quad (\mu_{13} = 0)\, ,  \label{eq.M.Xi} \\
\bm{M}_\Lambda = \bm{a}^\dag\bm{a} + \bm{A}_{33} \quad (\mu_{12} = 0)\, , \label{eq.M.Lambda}\\
\bm{M}_V = \bm{a}^\dag\bm{a} + \bm{A}_{22} + \bm{A}_{33} \quad (\mu_{23} = 0) \, , \label{eq.M.V}
\end{eqnarray}
%\end{subequations}
% 
which may be written in general as 
\begin{equation}
\label{generalM}
\bm{M} = \bm{a}^\dag\bm{a} + \lambda_2 \, \bm{A}_{22} + \lambda_3 \, \bm{A}_{33}
\end{equation}
with $\lambda_i$ as in Table~\ref{t1}.

\begin{table}%[h]
\caption{Values of $\lambda_{i}$, $i=2,\,3$, which determine the constant of motion $\bm{M}$.}
\begin{center}
\begin{tabular}{|c|cc|}
\hline
Configuration&$\lambda_{2}$&$\lambda_{3}$\\
\hline  & &\\[-3mm]
$\Xi$&1&2\\
$\Lambda$&0&1\\
$V$&1&1\\
\hline
\end{tabular} 
\end{center}
\label{t1}
\end{table}
  
\begin{figure}
\begin{center}
\includegraphics[width=0.3\linewidth]{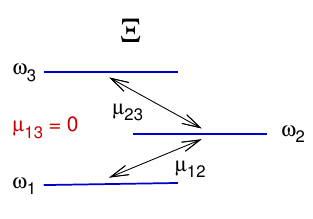} \quad \includegraphics[width=0.3\linewidth]{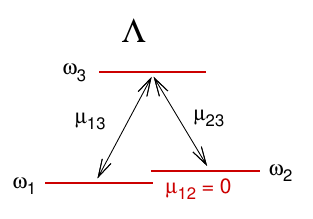} \quad \includegraphics[width=0.3\linewidth]{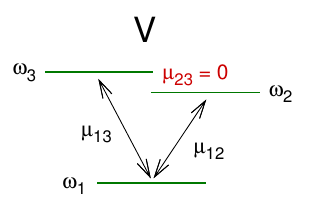}
\end{center}
\caption{Atomic configurations and dipolar coupling parameters.}
\label{fig0}
\end{figure}

The $\bm{M}$ operator corresponds to the total number of excitations for the different  atomic configurations $\Xi,\ \Lambda$ and ${\rm V}$ \cite{yoo85}.  The condition $\mu_{ij} = 0$ implies that transitions between levels $i$ and level $j$ are forbidden; a visual inspection of the different configurations (cf. Fig. \ref{fig0}) immediately suggests the expressions (\ref{eq.M.Xi}-\ref{eq.M.V}).
%Figura 0
\subsection{Semi-classical variational states}\label{GCS}

In the Hamiltonian that we have given above, for the description of three-level atoms interacting with an electromagnetic field, naturally appear matter operators that generate the unitary algebra in three dimensions, ${\rm U}(3)$. This lends itself to be described by the Gelfand-Tsetlin states~\cite{gelfand} which carry the irreducible representations of ${\rm U}(3)$ and are in general denoted by
\begin{eqnarray}\label{eq.G.state}
\left|\begin{array}{c c c c c} h_1 & & h_2 & & h_3 \\ & q_1 && q_2 & \\ && r && \end{array} \right\rangle = \vert h_1 \,  h_2 \,  h_3  \, q_1 \, q_2 \,  r \rangle \, ,
\end{eqnarray}   
where the labels satisfy the inequalities $q_1\geq r \geq q_2$ and $h_i\geq q_i\geq h_{i+1}$, with $i=1,2$.  The nine generators of $U(3)$ can be classified into weight, raising, and lowering operators. The weight generators $\bm{A}_{ii}$ satisfy the eigenvalue equations 
\begin{eqnarray*}
\bm{A}_{11} \vert h_1 \,  h_2 \,  h_3  \, q_1 \, q_2 \,  r \rangle = r \, \vert h_1 \,  h_2 \,  h_3  \, q_1 \, q_2 \,  r \rangle \, , \\
\bm{A}_{22} \vert h_1 \,  h_2 \,  h_3  \, q_1 \, q_2 \,  r \rangle = (q_1 +q_2 -r ) \, \vert h_1 \,  h_2 \,  h_3  \, q_1 \, q_2 \,  r \rangle \, , \\
\bm{A}_{33} \vert h_1 \,  h_2 \,  h_3  \, q_1 \, q_2 \,  r \rangle = (N_a - q_1 -q_2) \, \vert h_1 \,  h_2 \,  h_3  \, q_1 \, q_2 \,  r \rangle \, ,
\end{eqnarray*}
with $N_a=h_1+h_2 +h_3$.

For the values $q_1=h_1$, $q_2=h_2$, and $r=h_1$, one has the highest weight state, for which
\begin{equation}
\bm{A}_{ij} \vert h_1 \, h_2  \, h_3  \, h_1 \, h_2  \, h_1 \rangle =  0  \, , \quad \hbox{for} \ i < j   \, , 
\end{equation}
where $\bm{A}_{i j}$ are the raising weight generators. For this state, the eigenvalues of the weight generators determine the irreducible representation of $U(3)$, i.e., $[h_1,h_2,h_3]$. Physically this means that we have $h_i$ atoms in the level $\omega_i$.

In this work we may consider as a variational test function the direct product of a Heisenberg-Weyl coherent state (field contribution) with a $U(3)$ coherent state (matter contribution), because these generate a basis of the Hilbert space and let us obtain analytic expressions for the expectation values of matter and field observables, as done for two-level systems in~\cite{castanos09a}.

For the one-mode field we use the states $\vert\alpha\rangle$, which satisfy $\bm{a}|\alpha\rangle=\alpha|\alpha\rangle$, while for the matter we follow the procedure established by Perelomov~\cite{perelomov}.  The unnormalized ${\rm U}(3)$ coherent states can be defined as
\begin{equation}\label{eq.G}
|h_1h_2h_3,\vec{\gamma} \, \}= \bm{O}(\vec{\gamma})\vert h_1 \, h_2  \, h_3  \, h_1 \, h_2  \, h_1 \rangle \, ,
\end{equation}  
with $\vec{\gamma} = (\gamma_1,\gamma_2,\gamma_3)$, and where we have introduced the product of exponentials of lowering weight generators
\begin{eqnarray}
\bm{O}(\vec{\gamma}) &\equiv&{\rm e}^{\gamma_3\bm{A}_{21}}{\rm e}^{\gamma_2\bm{A}_{31}}{\rm e}^{\gamma_1\bm{A}_{32}} \, . 
\end{eqnarray}
Therefore, the variational test function is given by
	\begin{equation*}
	\vert h_1,h_2,h_3; \alpha\,\vec{\gamma}\rangle \equiv |\alpha\rangle\otimes|h_1,h_2,h_3;\vec{\gamma}\rangle\ .
	\end{equation*}
 
For the evaluation of the energy surface of the system, i.e., the expectation value of the Hamiltonian (\ref{eq.H.3level}) with respect to the tensorial product  $|\alpha\rangle\otimes|h_1,h_2,h_3;\vec{\gamma}\rangle$, we proceed as follows:
\begin{enumerate}
\item[i)] Determine the coherent state representations of the generators $\bm{A}_{ij}$, $\bm{a}$, and $\bm{a}^\dagger$.
\item[ii)] Evaluate the kernel of the Heisenberg-Weyl and ${\rm U}(3)$ coherent states, 
\begin{equation*}
\{ \alpha \, \vert \, \alpha^\prime \} \ \hbox{and}  \ \{ h_1h_2h_3,\vec{\gamma} \, \vert h_1h_2h_3,\vec{\gamma}^{\, \prime} \,  \} \
\end{equation*}
\item[iii)] Apply the representation form of each operator of the Hamiltonian to the corresponding kernel evaluated at $\alpha^\prime=\alpha$ and $\vec{\gamma}^{\, \prime}=\vec{\gamma}$.
\end{enumerate}

For the Heisenberg-Weyl case, it is well known that
\begin{equation*}
 \bm{a}  \to \frac{\partial \phantom{\alpha^*} }{\partial \alpha^*}  \, , \quad \bm{a}^\dagger \to \alpha^* \, , \quad
 \{ \alpha \, \vert \, \alpha^\prime \} = \exp{ (\alpha^* \, \alpha^\prime) } \, . 
\end{equation*}

Now, for the ${\rm U}(3)$ case, the first step is to determine the coherent state representation of the generators  
\begin{eqnarray}\label{eq.bAijk}
\{ h_1h_2h_3;\vec{\gamma}|\bm{A}_{ij}| \psi\rangle,
\end{eqnarray}
where $|\psi\rangle$ is an arbitrary state of the matter. Substituting the definition of the coherent state (\ref{eq.G}),  one 
has
\begin{eqnarray}\label{eq.bGijk}
\langle h_1 \ h_2 \, h_3 \, h_1 \, h_2 \, h_1 \vert  \bm{G}_{ij}\bm{O}^\dag(\vec{\gamma})|\psi\rangle,
\end{eqnarray}   
where we define $\bm{G}_{ij} = \bm{O}^\dag(\vec{\gamma})\bm{A}_{ij}{\bm{O}^\dag}^{-1}(\vec{\gamma})$.
Using the expansion of ${\rm e}^{\bm{A}} \bm{B}  {\rm e}^{-\bm{A}}$, it is straightforward that $\bm{G}_{ij} $ takes the form
\begin{eqnarray}\label{eq.Gij}
\bm{G}_{ij} &=&  \bm{A}_{ij} +  \delta_{3i}\gamma_1^* \bm{A}_{2j} +  \left[\delta_{1j}\left(\gamma_1^*\gamma_3^*- \gamma_2^*\right)-\gamma_1^*\delta_{2j}\right] \bm{A}_{i3}  \nonumber \\
&+&\delta_{i3}\left[\delta_{1j}\left({\gamma_1^*}^2\gamma_3^*- \gamma_1^*\gamma_2^*\right) - \delta_{2j}{\gamma_1^*}^2 \right]\bm{A}_{23}  \nonumber \\
&+& \left(\delta_{i3}\gamma_2^*+\delta_{2i}\gamma_3^*\right) \bm{A}_{1j} -\delta_{1j}\gamma_3^* \bm{A}_{i2} - \delta_{i3}\delta_{1j}\gamma_1^*\gamma_3^* \bm{A}_{22}  \nonumber \\
&+& \left\{\delta_{i3}\left[\delta_{1j}\left(\gamma_1^*\gamma_2^*\gamma_3^*- {\gamma_2^*}^2\right)-\delta_{2j}\gamma_1^*\gamma_2^*\right]  \right. \nonumber \\ 
&+& \left.\delta_{i2}\left[\delta_{1j}\left(\gamma_1^*{\gamma_3^*}^2- \gamma_2^*\gamma_3^*\right)-\delta_{2j}\gamma_1^*\gamma_3^*\right]\right\} \bm{A}_{13} \nonumber \\
&-& \delta_{1j}\left(\delta_{i3}\gamma_2^*\gamma_3^* + \delta_{i2}{\gamma_3^*}^2\right) \bm{A}_{12}. 
\end{eqnarray}

To apply $\bm{G}_{ij}$ to the bra associated to the highest weight state, we have to take into account that the weight generators are diagonal, the lowering generators yield zero, and the raising generators in (\ref{eq.Gij}) must be replaced by
\begin{eqnarray*}
\bm{A}_{23} \to \frac{\partial}{\partial \gamma_1^*}  \, , \quad \bm{A}_{13} \to \frac{\partial}{\partial \gamma_2^*}  \, , \quad
\bm{A}_{12} \to \left(\frac{\partial}{\partial \gamma_3^*}+\gamma_1^*\frac{\partial}{\partial \gamma_2^*}\right) \, . 
\end{eqnarray*}
This yields the Gelfand-Tsetlin coherent representation of the ${\rm U}(3)$ generators as 
\begin{equation*}
\mathcal{A}_{ij} \{h\,\vec{\gamma}\vert\psi\rangle = \{h\,\vec{\gamma}\vert\bm{A}_{ij}\vert\psi\rangle \, .
\end{equation*}

As an example, we give the ${\rm U}(3)$ coherent state representation of the ${\rm U}(2)$ subalgebra $\{ \bm{A}_{11},\, \bm{A}_{12},\, \bm{A}_{21},\, \bm{A}_{22}\}$: One writes, using (\ref{eq.Gij}),
\begin{eqnarray*}
\bm{G}_{11} &=& \bm{A}_{11} - \gamma_3^\ast\bm{A}_{12} + \left(\gamma_1^\ast \gamma_3^\ast - \gamma_2^\ast\right) \bm{A}_{13} \\
\bm{G}_{12} &=& \bm{A}_{12} - \gamma_1^\ast\bm{A}_{13} \\
\bm{G}_{21} &=& \bm{A}_{21} - \gamma_3^\ast\bm{A}_{12} + \gamma_3^\ast \left(\gamma_1^\ast \gamma_3^\ast - \gamma_2^\ast\right) \bm{A}_{13} \\
& +& \left(\gamma_1^\ast \gamma_3^\ast - \gamma_2^\ast\right) \bm{A}_{23} + \gamma_3^\ast(op{A}_{11} - \bm{A}_{22}) \\
\bm{G}_{22} &=& \bm{A}_{22} + \gamma_3^\ast\bm{A}_{12} - \gamma_1^\ast\gamma_3^\ast\bm{A}_{13} - \gamma_1^\ast\bm{A}_{23}
\end{eqnarray*}
Then we make the replacements indicated above, to get
\begin{eqnarray*}
\mathcal{A}_{11} &\to& h_1 - \gamma_2^\ast \frac{\partial}{\partial\gamma_2^\ast} - \gamma_3^\ast \frac{\partial}{\partial\gamma_3^\ast} \\
\mathcal{A}_{12} &\to& \frac{\partial}{\partial\gamma_3^\ast} \\
\mathcal{A}_{21} &\to& \gamma_3^\ast \left( h_1 - h_3 + \gamma_1^\ast \frac{\partial}{\partial\gamma_1^\ast} - \gamma_2^\ast \frac{\partial}{\partial\gamma_2^\ast} - \gamma_3^\ast \frac{\partial}{\partial\gamma_3^\ast} \right)  - \gamma_2^\ast \frac{\partial}{\partial\gamma_1^\ast} \\
\mathcal{A}_{22} &\to& h_2 - \gamma_1^\ast \frac{\partial}{\partial\gamma_1^\ast} + \gamma_3^\ast \frac{\partial}{\partial\gamma_3^\ast}
\end{eqnarray*}
It is straightforward to prove that the operators $\mathcal{A}_{ij}\ (i,j=1,2)$ satisfy the commutation relations of a ${\rm U}(2)$ algebra.

The ${\rm U}(3)$ matter kernel is given by
\begin{equation*}
 \langle  h_1 \, h_2  \, h_3  \, h_1 \, h_2  \, h_1 \vert \bm{O}^\dagger (\vec{\gamma}) \bm{O}(\vec{\gamma}^{\, \prime})\vert h_1 \, h_2  \, h_3  \, h_1 \, h_2  \, h_1 \rangle  \, .
\end{equation*}
To evaluate the expression it is convenient to rewrite the product of operators as
\begin{eqnarray}\label{eq.OdO}
\bm{O}^\dag(\vec{\gamma}) \, \bm{O}(\vec{\gamma}')=\bm{O}(\vec{\beta}) \, {\rm e}^{\lambda_1\bm{A}_{11} + \lambda_2\bm{A}_{22} + \lambda_3 \, \bm{A}_{33}}\bm{O}^\dag(\vec{\beta}') \, , \quad
\end{eqnarray}
because the matrix element, with respect to the Gelfand-Tsetlin highest weight  state, of the operators $\bm{O}(\vec{\beta})$ and $\bm{O}^\dag(\vec{\beta}' )$ yield a result equivalent to the identity operator and the remaining exponential is diagonal. 

To interchange the exponential operators, we use a faithful realization of the generators as $\bm{A}_{ij} = |i\rangle\langle  j|$. One then finds the values of the $\beta$'s, $\beta'$'s and $\lambda$'s as functions of $\gamma$'s and $\gamma'$'s, in such a manner that the expression (\ref{eq.OdO}) is satisfied. Following this procedure, one obtains the ${\rm U}(3)$ matter kernel
%\begin{widetext} 
\begin{eqnarray}\label{eq.Kp}
&& \{ h_1h_2h_3,\vec{\gamma} \, \vert h_1h_2h_3,\vec{\gamma}^{\, \prime} \,  \}  = \left(1+\gamma_2^*\gamma_2' + \gamma_3^*\gamma_3'\right)^{h_1-h_2} \times \nonumber\\ &&  
 \Big(1 + \gamma_2^* (\gamma^\prime_2 -\gamma^\prime_1 \, \gamma^\prime_3) + \gamma^*_1 ( \gamma^\prime_1 - \gamma^\prime_2 \, \gamma^*_3 + \gamma^\prime_1 \, \gamma^*_3 \, \gamma^\prime_3 ) \Big)^{h_2-h_3} \,. 
\end{eqnarray}
%\end{widetext}
The general case of distinguishable particles could be interesting in quantum information theory, for example, for the description of systems of q-trits. In our case of study we restrict ourselves to the totally symmetric configuration, i.e., that of indistinguishable particles. For the symmetric basis the corresponding kernel of the matter contribution is obtained by taking $h_2=h_3=0$. From here on we simplify the notation by omitting the values of $h_2$ and $h_3$ in the Gelfand-Tsetlin states. Therefore the kernel of the tensorial product of coherent states is
\begin{eqnarray}\label{eq.K}
K(h_1;\alpha,\alpha',\vec{\gamma},\vec{\gamma}') = {\rm e}^{\alpha^*\alpha'}\left(1+\gamma_2^*\gamma_2' + \gamma_3^*\gamma_3'\right)^{h_1}.
\end{eqnarray}

\subsection{Energy surface}\label{energy_surface}

Applying the corresponding coherent state representation of the Hamiltonian (\ref{eq.H.3level}) on the kernel above, dividing by the scalar product of the coherent states,  and replacing $\alpha'=\alpha$ and $\vec{\gamma}' = \vec{\gamma}$, the energy surface is
\begin{eqnarray}\label{eq.E.G}
{\cal E}^c &=& \Omega \rho^2 + h_1 \frac{\omega_1 +\omega_2\varrho_3^2 + \omega_3 \varrho_2^2}{1 + \varrho_2^2 + \varrho_3^2 } 
- 2\sqrt{h_1}\mu_{12} \frac{\rho\varrho_3\cos(\vartheta_{3})}{1 + \varrho_2^2 + \varrho_3^2 }\nonumber \\ &-& 2\sqrt{h_1}\mu_{13} \frac{\rho\varrho_2\cos(\vartheta_{2})}{1 + \varrho_2^2 + \varrho_3^2 }  - 2\sqrt{h_1}\mu_{23} \frac{\rho\varrho_2\varrho_3\cos(\vartheta_{1})}{1 + \varrho_2^2 + \varrho_3^2 },
\end{eqnarray}
where we have rewritten the parameters in their polar form, i.e., $\alpha = \rho {\rm e}^{i\phi},\ \gamma_j = \varrho_j {\rm e}^{i\varphi_j}$ and  identified $\vartheta_3 = \phi-\varphi_3, \ \vartheta_2 = \phi-\varphi_2$ and $\vartheta_1 = \phi -\varphi_2+\varphi_3$. 

Minimizing ${\cal E}^c$ respect to the phases $\vartheta_i$ one finds that the critical values are given by  $\vartheta_{ic} = 0,\ \pi$. The minimum is obtained when $\mu_{ij}\cos\left(\vartheta_{kc}\right)>0$ for cyclic indices $i,\ j$ and $k$. Since these values are independent of $\rho$'s, one may replace this condition on Eq. (\ref{eq.E.G}), and hence the energy surface is rewritten as
\begin{eqnarray}\label{eq.E.G2}
{\cal E}^c &=& \Omega \rho^2 + h_1 \frac{\omega_1 +\omega_2\varrho_3^2 + \omega_3 \varrho_2^2}{1 + \varrho_2^2 + \varrho_3^2 }  \nonumber \\
&-& 
2\sqrt{h_1}\rho \frac{|\mu_{12}|\varrho_3 + |\mu_{13}|\varrho_2 + |\mu_{23}|\varrho_2\varrho_3}{1 + \varrho_2^2 + \varrho_3^2 }. 
\end{eqnarray}
It is easy to see that the condition 
\begin{eqnarray}
\frac{\partial}{\partial \rho}{\cal E}^c =0,
\end{eqnarray}  
is satisfied when $\rho=\rho_c$ (critical value of the variable $\rho$) where $\rho_c$ is given by
\begin{eqnarray}\label{eq.rho.minG}
\rho_c = \frac{\sqrt{h_1}}{\Omega} \frac{|\mu_{12}|\varrho_{3c} + |\mu_{13}|\varrho_{2c} + |\mu_{23}|\varrho_{2c}\varrho_{3c}}{1 + \varrho_{2c}^2 + \varrho_{3c}^2} \, .
\end{eqnarray}
Here $\varrho_{2c}$ and $\varrho_{3c}$ stand for the critical values of $\varrho_{2}$ and $\varrho_{3}$, respectively.

It is worth stressing the fact that the energy surface given by Eq. (\ref{eq.E.G}) [or equivalently Eq. (\ref{eq.E.G2})] has no a dependence on $\gamma_1 = \varrho_1 {\rm e}^{i\varphi_1}$, because we are taking $h_2=h_3=0$ in the definition of the Gelfand-Tsetlin coherent state. 

For the semi-classical calculation of the ground state energy, it is worth referring to the intensive quantity $E^c = {\cal E}^c/ h_1$ which describes the energy per particle:
\begin{eqnarray}\label{eq.E.G3}
E^c &=&  \Omega \, r^2 +  \frac{\omega_1 +\omega_2 \varrho_3^2 + \omega_3 \varrho_2^2}{1 + \varrho_2^2 + \varrho_3^2 }  \nonumber \\
&-&
2r \frac{|\mu_{12}|\varrho_3 + |\mu_{13}|\varrho_2 + |\mu_{23}|\varrho_2\varrho_3}{1 + \varrho_2^2 + \varrho_3^2 }, 
\end{eqnarray}
where $r=\rho/\sqrt{h_1}$. In a similar way we define the total number of excitations per particle $M^c = {\cal M}^c/N_a$.  

An approximation to the ground state energy of the system is obtained by substituting the minima critical points into the energy surface. From (\ref{eq.rho.minG}) and (\ref{eq.E.G3}) we obtain $E^{c}=E^{c}(\varrho_{2c},\,\varrho_{3c})$, whose minimum in general has no analytic solutions for arbitrary points in parameter space ($\mu_{ij}$) and a particular atomic configuration.

The critical points satisfy $\varrho_{2c},\,\varrho_{3c}\geq0$. To find numerically these critical points we proceed as follows, starting with the first quadrant in the $\varrho_{2c}-\varrho_{3c}$ plane:
\begin{itemize}
\item The area is divided into $N$ regions forming a lattice; 
\item the energy surface is evaluated at the central point of each of these regions;
\item the region with minimum energy, together with its closest neighbors, is selected to build a new lattice;
\item this method is iterated until the desired precision is reached.
\end{itemize}
If the area of the first set is $S$, the method establishes the critical point with a precision of $3^{m-1}  \sqrt{S/N^m}$, where $m$ is the number of iterations.

Recently~\cite{clpcna13} we found the minimum energy surface $E^c$ as a function of the control parameters $\mu_{ij}$. It changes value from $E^c=0$ to  $E^c<0$, when a transition from $M^c=0$ (normal regime) to $M^c>0$ (collective regime) in the total number of excitations of the corresponding semi-classical approximation to the ground state of the system takes place.  This leads to the existence of a separatrix in parameter space, for which we were able to propose the following ansatz:

For the $\Xi$ configuration,
%
%\begin{subequations}
%
\begin{eqnarray}\label{eq.trans.Xi}
\Omega\, \omega_{21} = \mu_{12}^2 + \left[|\mu_{23}|-\sqrt{\Omega \, \omega_{31}}\right]^2 \Theta\left[|\mu_{23}|-\sqrt{\Omega \, \omega_{31}}\right], \qquad
\end{eqnarray}
where the Bohr frequency $\omega_{ij}\equiv \omega_i-\omega_j$ is the energy shift between the atomic levels $i$ and $j$  and  $\Theta\left[x\right]$ stands for the Heaviside theta function.

For the $\Lambda$ configuration,
\begin{eqnarray}\label{eq.trans.Lambda}
\Omega\,\omega_{31} =\mu_{13}^2 +  \left[|\mu_{23}|-\sqrt{\Omega \, \omega_{21}}\right]^2   \Theta\left[|\mu_{23}|-\sqrt{\Omega \, \omega_{21}}\right]\, ,
\end{eqnarray}

For the $V$ configuration,
\begin{eqnarray}\label{eq.trans.V}
\frac{\mu_{12}^2}{\Omega\,\omega_{21}} + \frac{\mu_{13}^2}{\Omega\,\omega_{31}}= 1.
\end{eqnarray}
%
%\end{subequations}
The separatrix of the different configurations correspond to the thermodynamic limit, that is when the number of atoms $N_a \to \infty$.
\subsection{Order of the transitions}\label{order.trans}

A phase transition is of order $j$, according to the Ehrenfest classification~\cite{gilmore93}, if $j$ is the lowest non-negative integer for which 
\begin{equation*}
\lim_{\epsilon \to 0} \frac{\partial^j E^c}{\partial s^j} \Bigg |_{s=s_0+\epsilon} \neq \lim_{\epsilon \to 0} \frac{\partial^j E^c}{\partial s^j} \Bigg |_{s=s_0-\epsilon} \, ,
\end{equation*}
where $s$ represents here any of the control parameters $\mu_{ij}$.
In general we do not have analytical expressions for the critical points, so the order of the transitions must be obtained numerically. In the case of first-order transitions, however, we may use
\begin{eqnarray*}
dE^c = \ &&\left(\frac{\partial E^c}{\partial \rho}\right) d\rho + \left(\frac{\partial E^c}{\partial \varrho_2}\right) d\varrho_2 + \left(\frac{\partial E^c}{\partial \varrho_3}\right) d\varrho_3 + \sum_{i<j} \left(\frac{\partial E^c}{\partial \mu_{ij}}\right) d\mu_{ij} 
\end{eqnarray*}
which evaluated at the critical points reduces to
\begin{equation*}
dE^c \Bigg|_{\rho_c,\varrho_{2c},\varrho_{3c}} = \ \sum_{i<j} \left(\frac{\partial E^c}{\partial \mu_{ij}}\right)_{\rho_c,\varrho_{2c},\varrho_{3c}} d\mu_{ij} 
\end{equation*}
and this provides us with the following expressions:

\noindent For the $\Xi$ configuration
%\begin{subequations}
\begin{eqnarray}\label{eq.dE.X}
\frac{\partial}{\partial \mu_{12}}  E^c_{\Xi} = - 2  \frac{r_c\ \varrho_{3c}}{1+\varrho_{2c}^2 + \varrho_{3c}^2}, \\
\frac{\partial}{\partial \mu_{23}} E^c_{\Xi} = - 2  \frac{r_c\ \varrho_{2c}\ \varrho_{3c}}{1+\varrho_{2c}^2 + \varrho_{3c}^2}\,;
\end{eqnarray}  
%\end{subequations}
%
for the $\Lambda$ configuration
%
%\begin{subequations}
\begin{eqnarray}\label{eq.dE.L}
\frac{\partial}{\partial \mu_{13}}E^c_{\Lambda} = - 2  \frac{r_c\ \varrho_{2c}}{1+\varrho_{2c}^2 + \varrho_{3c}^2}, \\
\frac{\partial}{\partial \mu_{23}} E^c_{\Lambda} = - 2  \frac{r_c\ \varrho_{2c}\ \varrho_{3c}}{1+\varrho_{2c}^2 + \varrho_{3c}^2}\,;
\end{eqnarray}  
%\end{subequations}
%
and for the ${\rm V}$ configuration
%
%\begin{subequations}
\begin{eqnarray}\label{eq.dE.V}
\frac{\partial}{\partial \mu_{12}} E^c_{V} = - 2  \frac{r_c\ \varrho_{3c}}{1+\varrho_{2c}^2 + \varrho_{3c}^2}, \\
\frac{\partial}{\partial \mu_{13}} E^c_{V} = - 2  \frac{r_c\ \varrho_{2c}}{1+\varrho_{2c}^2 + \varrho_{3c}^2}\,.
\end{eqnarray}  
%\end{subequations}

For the second-order transitions one has to infer them through numerical differentiation of the equations, or through derivatives of second order when analytical expressions are available.

\subsection{Statistics of semi-classical ground state}\label{fluc}

The statistics of the semi-classical ground state is given by the well-known $Q$-Mandel parameter \cite{mandel79}, defined for the field states as
\begin{eqnarray}
Q = \frac{(\Delta n)^2 - \langle  \bm{n}\rangle}{\langle  \bm{n}\rangle}.
\end{eqnarray}
The photon distribution obeys $(\Delta n)^2 = \langle {\bf n}\rangle$, and hence $Q=0$ for any value of the control parameters, i.e., the contribution of the photons in the semi-classical ground state obeys Poissonian statistics.    

On the other hand, one may study the statistics of the ground state as a function of the total number of excitations ${\cal M}$, i.e., consider both field and matter contributions. So one may define, in a similar way, the $Q_M$-Mandel parameter as   
\begin{eqnarray}\label{eq.QM.def}
Q_M = \frac{(\Delta M)^2 - \langle  \bm{M}\rangle}{\langle  \bm{M}\rangle}.
\end{eqnarray}

To evaluate the expression (\ref{eq.QM.def}) we use Eq.(\ref{generalM}) together with
\begin{eqnarray}
\bm{M}^2 &=& \bm{n}^2 + \lambda_2^2\bm{A}_{22}^2 + \lambda_3^2\bm{A}_{33}^2  \nonumber \\
&+& 2\bm{n}\left(\lambda_2\bm{A}_{22} + \lambda_3\bm{A}_{33}\right) + 2\lambda_2\lambda_3\bm{A}_{22}\bm{A}_{33} \, .
\end{eqnarray}
For the totally symmetric coherent variational test function one may establish the following relations between expectation values for matter and field observables:
%\begin{subequations}
\begin{eqnarray}\label{eq.relations.2}
\langle  \bm{n}^2\rangle = \langle  \bm{n}\rangle^2 + \langle \bm{n}\rangle,\\
\langle  \bm{A}_{22}^2\rangle = \langle  \bm{A}_{22}\rangle +\left(1-\frac{1}{N_a}\right) \langle \bm{A}_{22}\rangle^2,\\
\langle  \bm{A}_{33}^2\rangle = \langle  \bm{A}_{33}\rangle +\left(1-\frac{1}{N_a}\right) \langle \bm{A}_{33}\rangle^2,\\
\langle  \bm{n}\bm{A}_{ii}\rangle = \langle  \bm{n}\rangle \langle \bm{A}_{ii}\rangle, \\
\langle  \bm{A}_{22}\bm{A}_{33}\rangle = \left(1-\frac{1}{N_a}\right)\langle  \bm{A}_{22}\rangle \langle \bm{A}_{33}\rangle \,.
\end{eqnarray} 
%\end{subequations}    
%
%***************************************************
%Figura 1
\begin{figure}
\begin{center}
\includegraphics[width=0.48\linewidth]{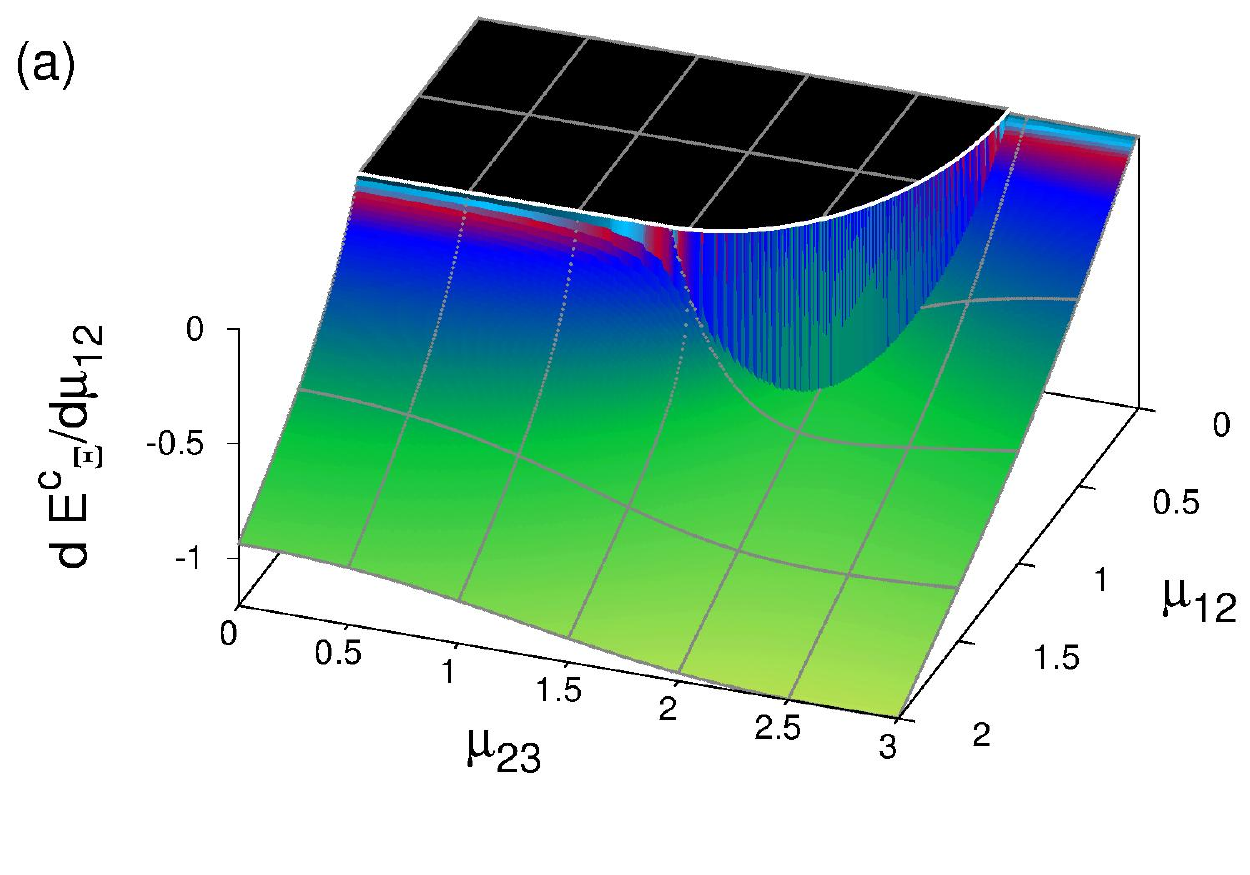}\,
\includegraphics[width=0.48\linewidth]{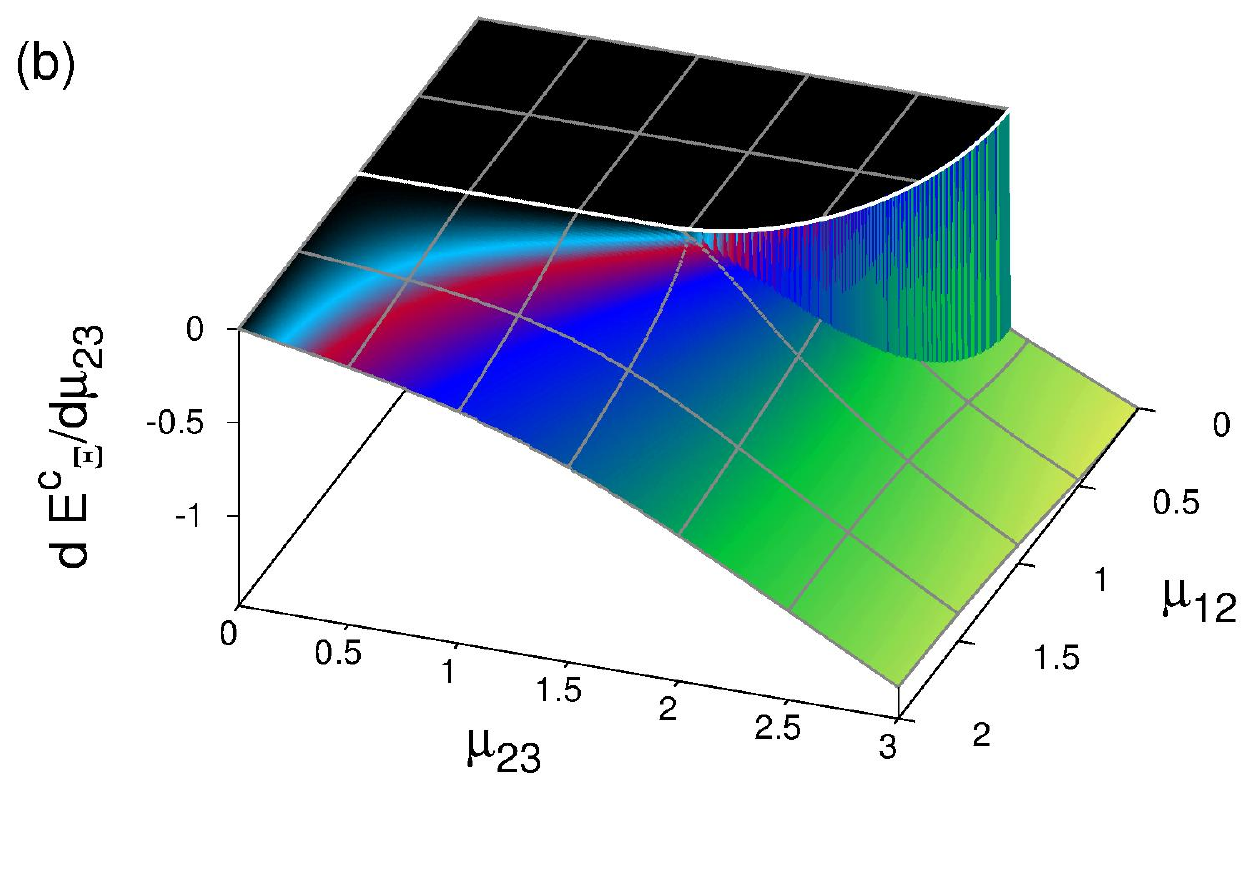}
\end{center}
\caption{(Color online.)  First derivative of the ground state energy with respect its control parameters, for atoms in $\Xi$ configuration under double resonance condition $\Delta_{21}=\Delta_{32}= \nobreak 0$: (a) with respect to $\mu_{12}$, and (b) with respect to $\mu_{23}$.}\label{f1}
\end{figure}
% 
%Figura 2
\begin{figure*}
\begin{center}
\includegraphics[width=0.48\linewidth]{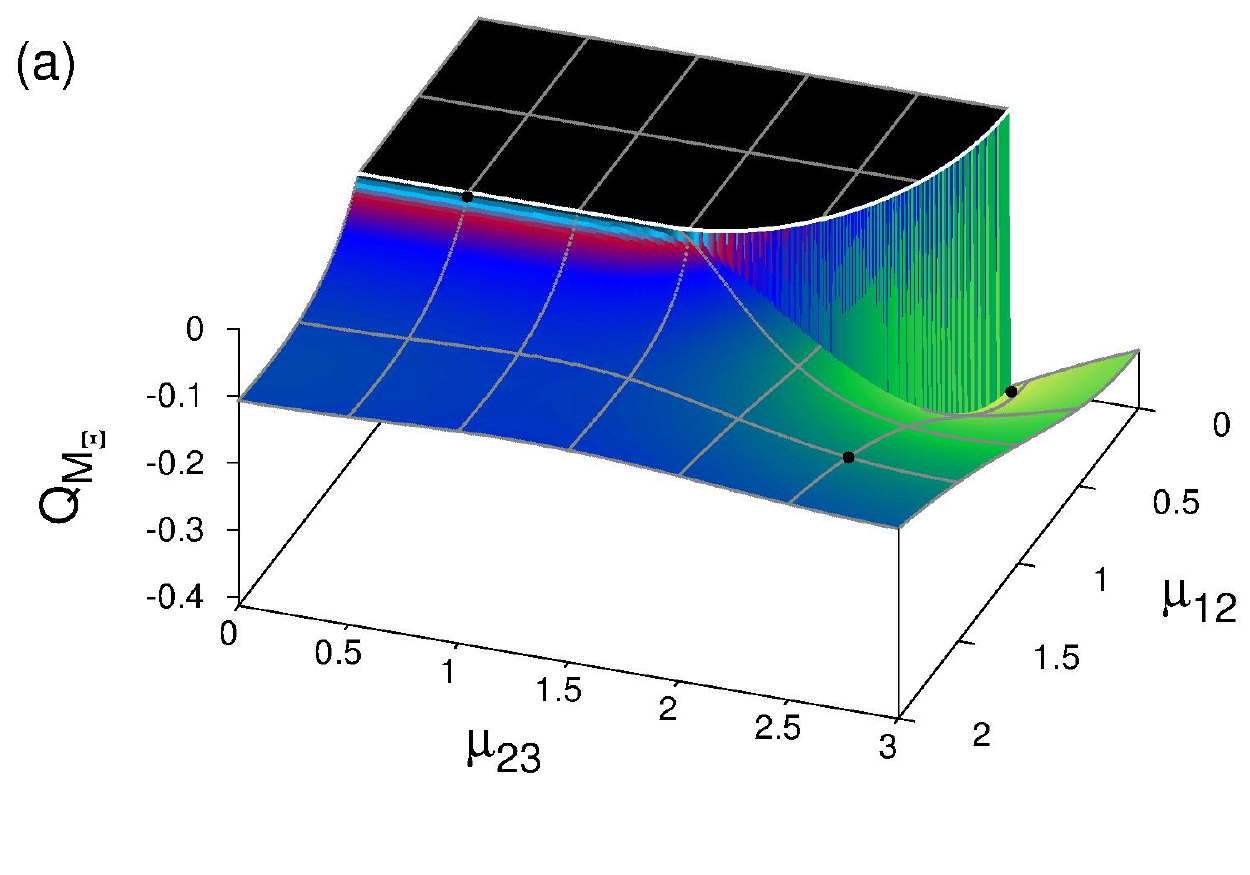}\
\includegraphics[width=0.48\linewidth]{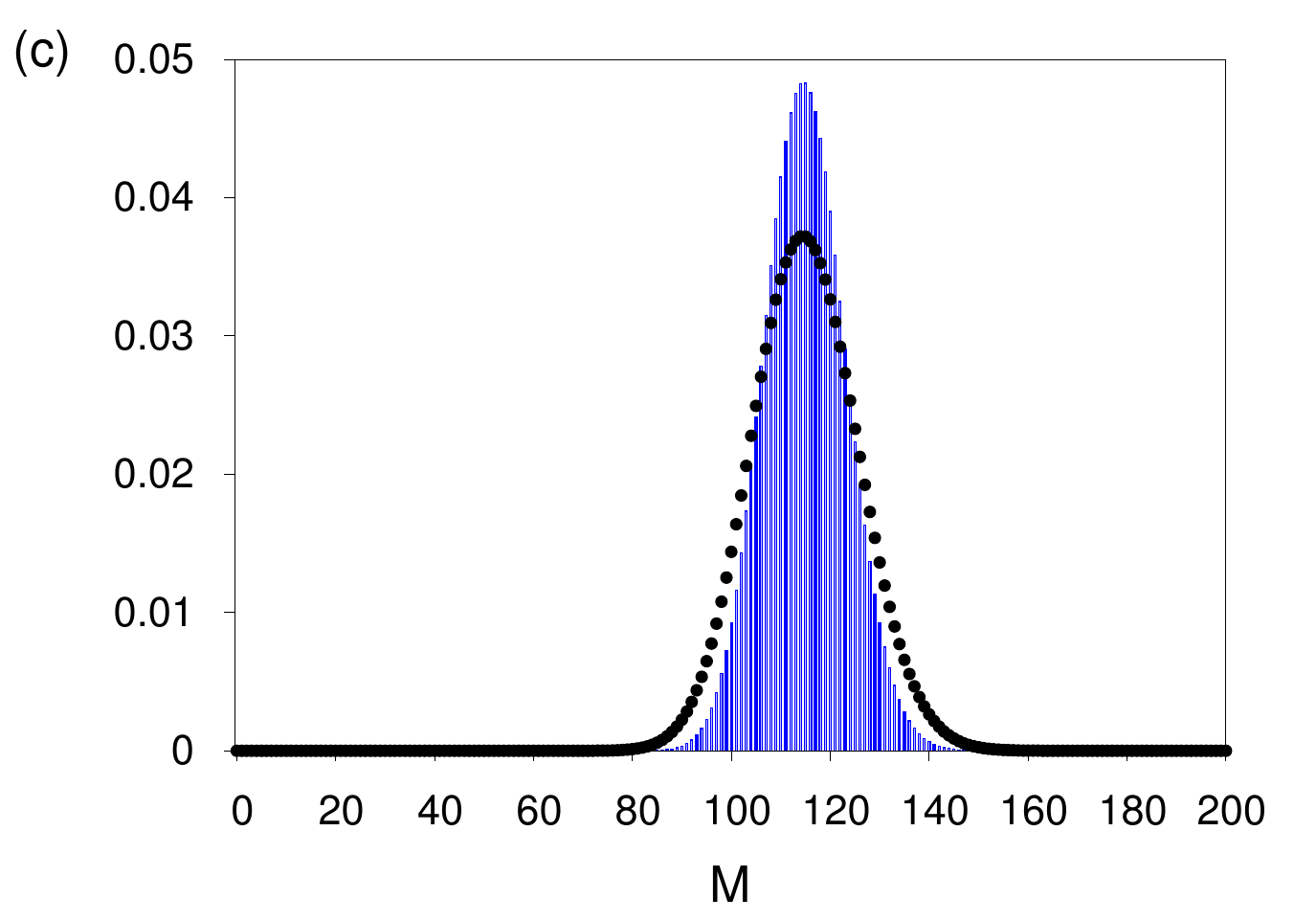}\\
\includegraphics[width=0.48\linewidth]{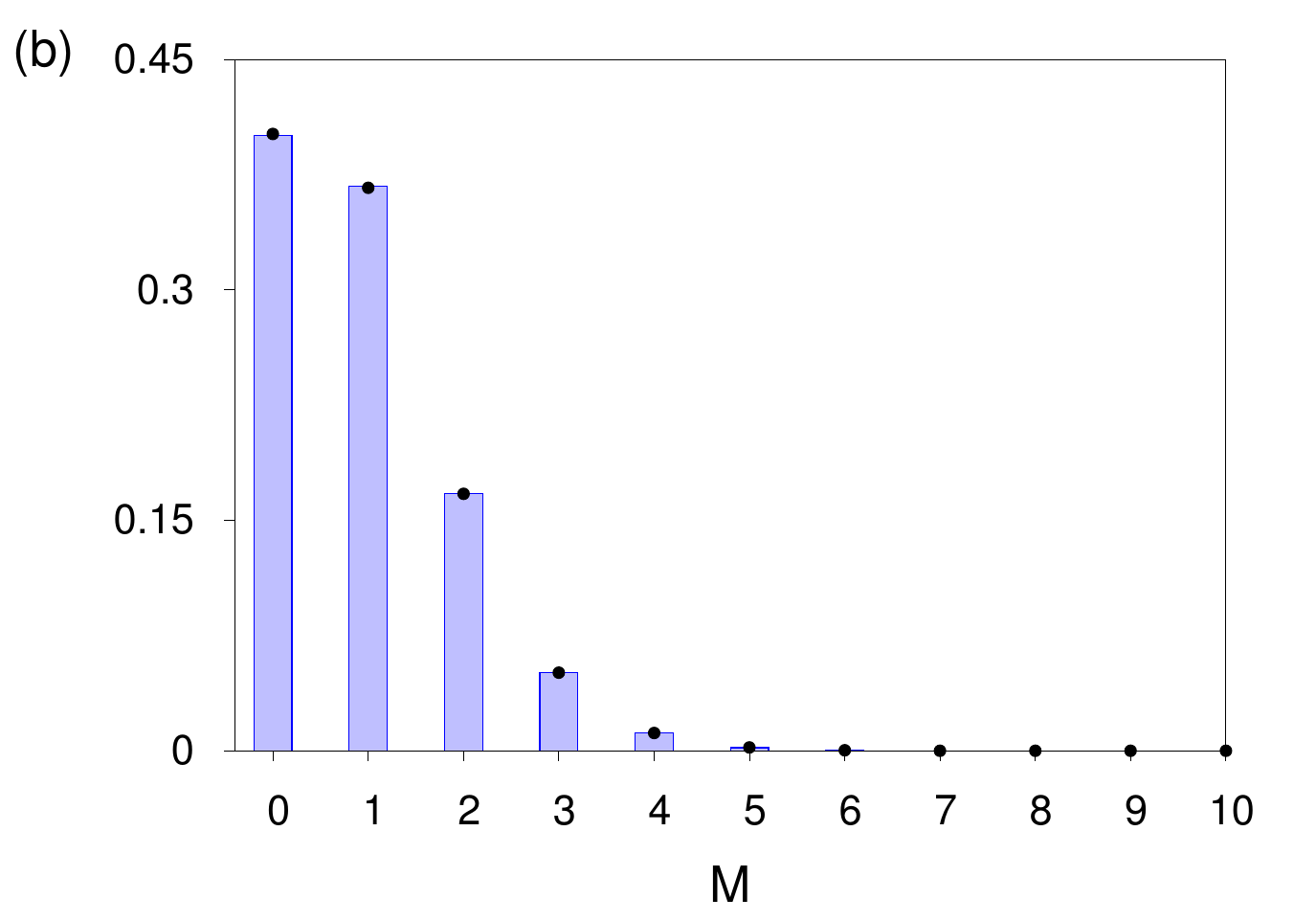}\
\includegraphics[width=0.48\linewidth]{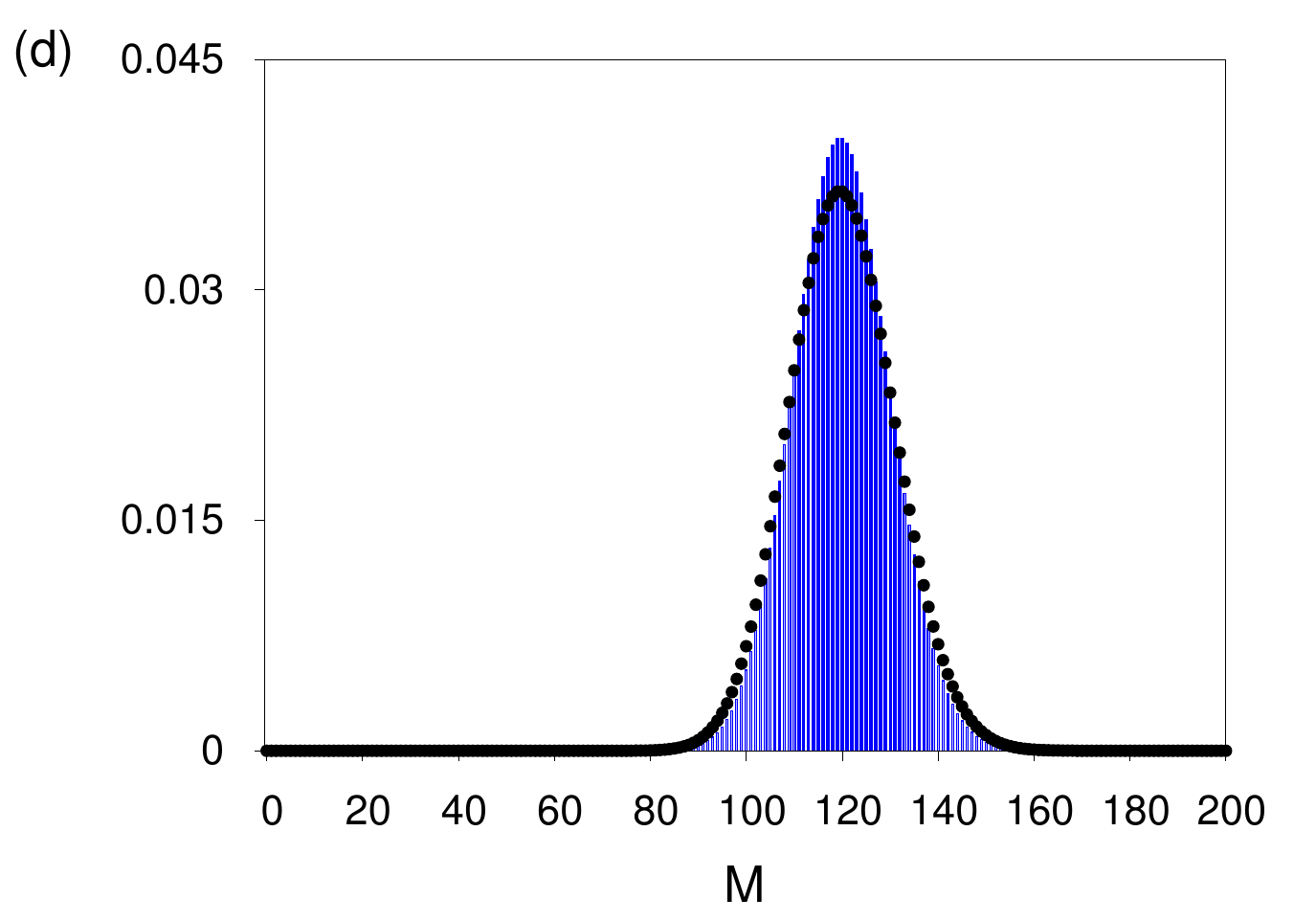}
\end{center}
\caption{(Color online.) (a) $Q_M$-Mandel parameter as a function of the control parameters, for atoms in $\Xi$ configuration in double resonance. The separatrix is shown by a white line, as well as three points (dots) where the corresponding $M$ distribution of the ground state for $N_a=40$ atom has been calculated (solid bars) and compared with its corresponding Poissonian distribution (dots). (b) ${\cal M}$ distribution with parameters $\mu_{12} = 1.01,\ \mu_{23}= 0.5$ providing values $M^c_\Xi \approx  2.3 \times 10^{-2}$ and $Q_M \approx -4.7\times10^{-3}$. (c) ${\cal M}$ distribution with parameters $\mu_{12} = 0.05,\ \mu_{23}= 2.45,$ providing values $ M^c_\Xi \approx 2.87 $ and $Q_M\approx -0.41$. (d) ${\cal M}$ distribution with $\mu_{12} = 1.5,\ \mu_{23}=2.5,$ providing values $M^c_\Xi \approx 3$ and $Q_M\approx -0.17$.}\label{f2}
\end{figure*}
%
%Figura 3
\begin{figure}
\begin{center}
\includegraphics[width=0.7\linewidth]{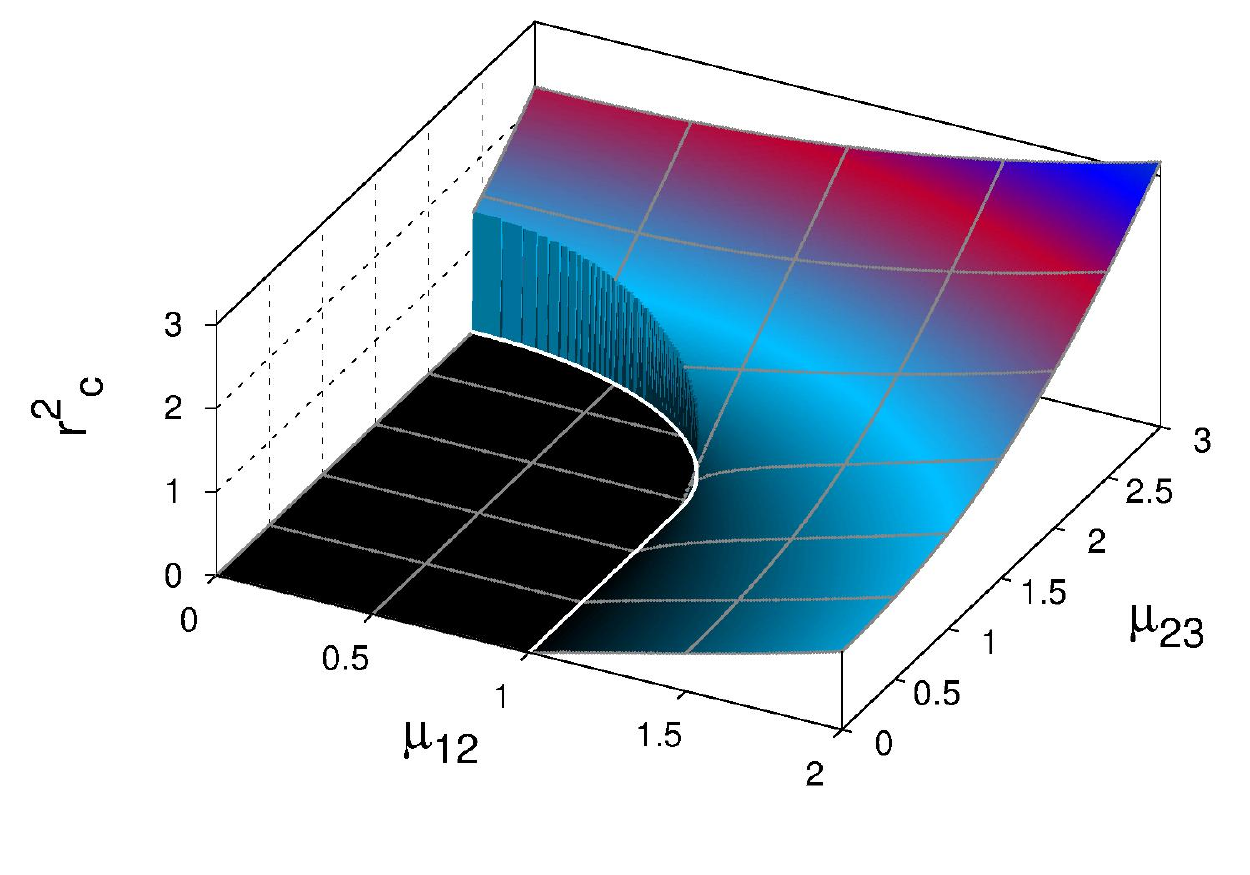}
\end{center}
\caption{(Color online.) The average number of photons in units of the total number of atoms $r^2_c = \rho^2_c/N_a$ is shown, for atoms in $\Xi$ configuration in double resonance. Notice that for greater values of $\mu_{23}=\sqrt 2$, along the separatrix there are coexistence between two values of number of photons.}\label{f3}
\end{figure}
%
%*******************************************************

\noindent Hence, the fluctuation of the total number of excitations for the variational state, defined by $(\Delta M^c)^2 = \langle  \bm{M}^2\rangle - \langle \bm{M}\rangle^2$, is given by
\begin{eqnarray}\label{eq.fluc.MG}
(\Delta M^c)^2 = \langle  \bm{M}\rangle + \lambda_3\left(\lambda_3-1\right)\langle  \bm{A}_{33}\rangle -\frac{1}{N_a}\left[\lambda_2\langle \bm{A}_{22}\rangle + \lambda_3\langle \bm{A}_{33}\rangle\right]^2,
\end{eqnarray}
where we have used the fact that $\lambda_2^2=\lambda_2$ and $\lambda_3^2 = \lambda_3$ or $2\lambda_3$ to identify the appropriate value of $\langle \bm{M}\rangle$. Then the $Q_M$-Mandel parameter for this state reads
\begin{eqnarray}\label{eq.QM}
Q_M{}^c = \frac{1}{\langle  \bm{M}\rangle} \bigg[\lambda_3\left(\lambda_3-1\right)\langle  \bm{A}_{33}\rangle -\frac{1}{N_a}\left[\lambda_2\langle \bm{A}_{22}\rangle + \lambda_3\langle \bm{A}_{33}\rangle\right]^2\bigg].  
\end{eqnarray}
Note that the $Q_M$-Mandel parameter does not depend on the total number of atoms $N_a$, since both quantities $\langle  \bm{M}\rangle$ and $\langle  \bm{A}_{ii}\rangle$ are proportional to $N_a$ 

Since $\lambda_3=1$ for the $\Lambda$ and ${\rm V}$ configurations,  one finds from Eq. (\ref{eq.QM}) that in these cases $Q_M\leq 0$, and then the corresponding coherent state obeys only Poissonian ($Q_M=0$) and sub-Poissonian ($Q_M<0$) statistics. For the $\Xi$ configuration however $\lambda_3=2$ and hence the sign of $Q_M$ may be determined only via evaluation of  the corresponding critical points.

\subsection{Numerical results}\label{numerical.results}

As pointed out in \cite{clpcna13}, the minimization of the semi-classical energy $E^c$ provides analytic expressions for the  phases and $r_c=\rho_c/\sqrt{N_a}$. There is not, in general, an analytic solution available for the minimum value of the energy surface with respect to the other two independent variables $\varrho_2$ and $\varrho_3$. This suggests the use of a numerical method to evaluate the critical points $\varrho_{2c},\ \varrho_{3c}$, as functions of the control parameters $\mu_{ij}$.
 
To describe the levels of the atom we can use the detuning, defined by
\begin{eqnarray}\label{eq.detuning}
\Delta_{ij} = \omega_{ij} - \Omega, \quad \omega_{ij} = \omega_i-\omega_j\,.
\end{eqnarray}
Without loss of generality, we chose $\Omega=1$ and $\omega_1 = 0$. So both the control parameters, atomic levels and the detuning are measured in units of the field frequency.

\subsubsection{$\Xi$ configuration}\label{NR.Xi}
The $\Xi$ configuration forbids the transition $\omega_1\longleftrightarrow\omega_3$, and this is introduced in the Hamiltonian by taking $\mu_{13}=\nobreak0$. Then $\Delta_{21}$ and $\Delta_{32}$ are related to the energy levels by
%
%\begin{subequations}
\begin{eqnarray}
\omega_2 =  \Delta_{21} + \omega_1 + \Omega\,,\\
\omega_3 = \Delta_{32} +  \Delta_{21} + \omega_1 + 2\,\Omega\,.
\end{eqnarray}
%\end{subequations}

Also, in the $\Xi$ configuration the condition $\omega_2\approx \omega_3/2$ is fulfilled, and the detuning should satisfy $\Delta_{21}\approx \Delta_{32}$ with $|\Delta_{ij}|< 1$ to be consistent with the RWA approximation. 

Figure \ref{f1} shows the first derivatives of the energy surface for the ground state in double resonance, i.e., when $\Delta_{21}=\Delta_{32}=0$.  The corresponding separatrix Eq. (\ref{eq.trans.Xi}) is shown by a white line. One can observe that the derivative is continuous in the region $\mu_{23}\leq \sqrt{\Omega \, \omega_{31}}$, where the separatrix is given by $\mu_{12}= \sqrt{\Omega \, \omega_{21}}$; here a second-order transition occurs. For $\mu_{23}> \sqrt{\Omega \, \omega_{31}}$, the separatrix is given by $\left(|\mu_{23}|- \sqrt{\Omega \, \omega_{31}}\right)^2+{\mu_{12}}^2 = \Omega \, \omega_{21}$, the derivative is discontinuous and first-order transitions take place.   

In Fig. \ref{f2}(a) the $Q_M$-Mandel parameter Eq. (\ref{eq.QM}) for this configuration is shown. One can observe that for
\begin{eqnarray}\label{eq.Xi.Poisson}
\Omega\,\omega_{21} &\geq& \mu_{12}^2 + \left(|\mu_{23}|-\sqrt{\Omega\,\omega_{31}}\right)^2 \Theta\left[|\mu_{23}|-\sqrt{\Omega\,\omega_{31}}\right]
\end{eqnarray}
we have $Q_M=0$, i.e., for this region in parameter space the semi-classical ground state has Poissonian statistics. On the other hand, when Eq. (\ref{eq.Xi.Poisson}) is not satisfied one finds $Q_M<0$, providing sub-Poissonian statistics. 

Figures \ref{f2}(b), (c) and (d), respectively, show the distribution of ${\cal M}$ in the semi-classical ground state for $N_a=40$ atoms, using $(\mu_{12} = 1.01,\ \mu_{23}= 0.5)$,\ $(\mu_{12} = 0.05,\ \mu_{23}= 2.45)$ and $(\mu_{12} = 1.5,\ \mu_{23}=2.5)$ (solid bars) in comparison with the corresponding Poissonian distribution (dots). Notice that the first two points are very close to the separatrix, as shown in the figure \ref{f2}(a) [dotted data], but their corresponding $Q_M$-Mandel parameters are different. One may observe that in Fig. \ref{f2}(b) both distributions are practically indistinguishable, since the average value of the total number of excitations is $M^c_\Xi \approx  2.3 \times 10^{-2}$ (with the corresponding $Q_M$-Mandel parameter $Q_M \approx -4.7\times10^{-3}$), i.e., the contribution of the state with ${\cal M}=0$ dominates in the ground state, while, in the other cases [Figs. \ref{f2}(c), (d)]  the average values are $M^c_\Xi \approx 2.87 \hbox{ and } 3$ (with $Q_M\approx-0.41 \hbox{ and } -0.17$, respectively), where the contribution of the state with ${\cal M}=0$ is negligible. 

Fig. \ref{f3} shows the average number of photons in units of the total number of atoms, $r_c^2 = \rho_c^2/N_a = \langle  \bm{n}\rangle/N_a$. Since the field is a coherent state, the fluctuation of the number of photons satisfies $(\Delta n)^2=\langle \bm{n}\rangle$.

\subsubsection{$\Lambda$ configuration}\label{NR.Lambda}
%

%Figura 4
\begin{figure}
\begin{center}
\includegraphics[width=0.48\linewidth]{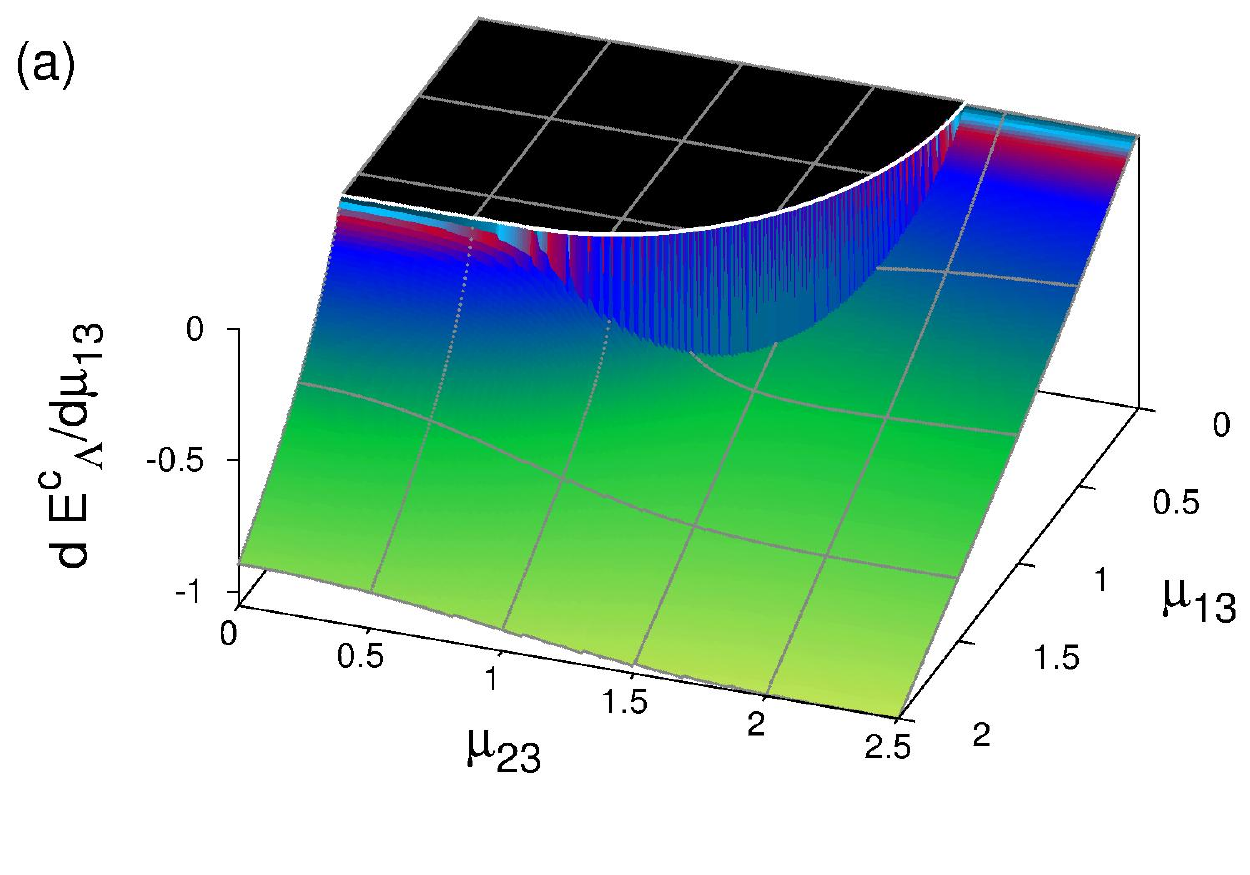}\,
\includegraphics[width=0.48\linewidth]{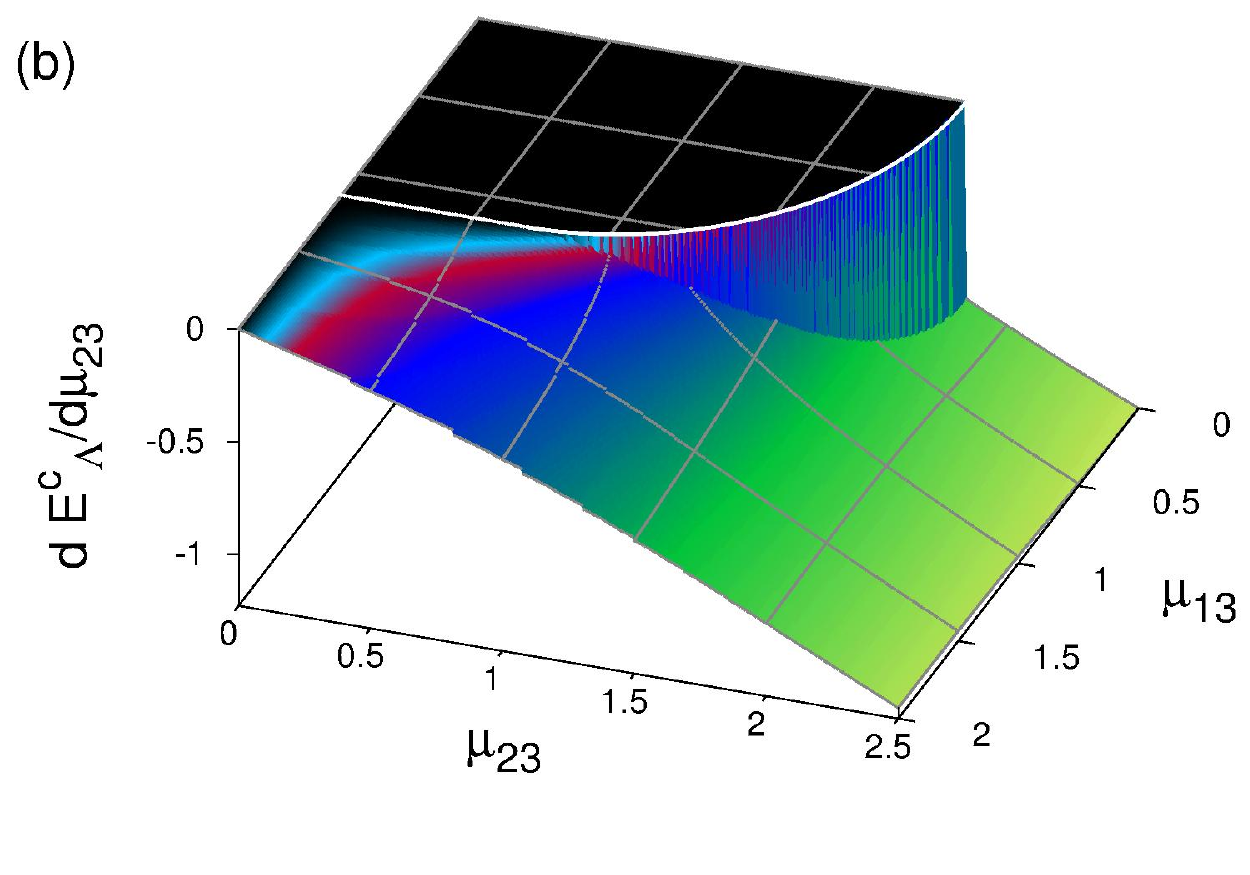}
\end{center}
\caption{(Color online.) First derivative of the ground state energy with respect to its control parameters, for atoms in the $\Lambda$ configuration with a non-resonant condition $\Delta_{31}=0.3$ and $\Delta_{32}=-0.2$. (a) Derivative with respect to $\mu_{13}$, and (b) derivative with respect to $\mu_{23}$. }\label{f4}
\end{figure}
% Figura 5
\begin{figure*}
\begin{center}
\includegraphics[width=0.48\linewidth]{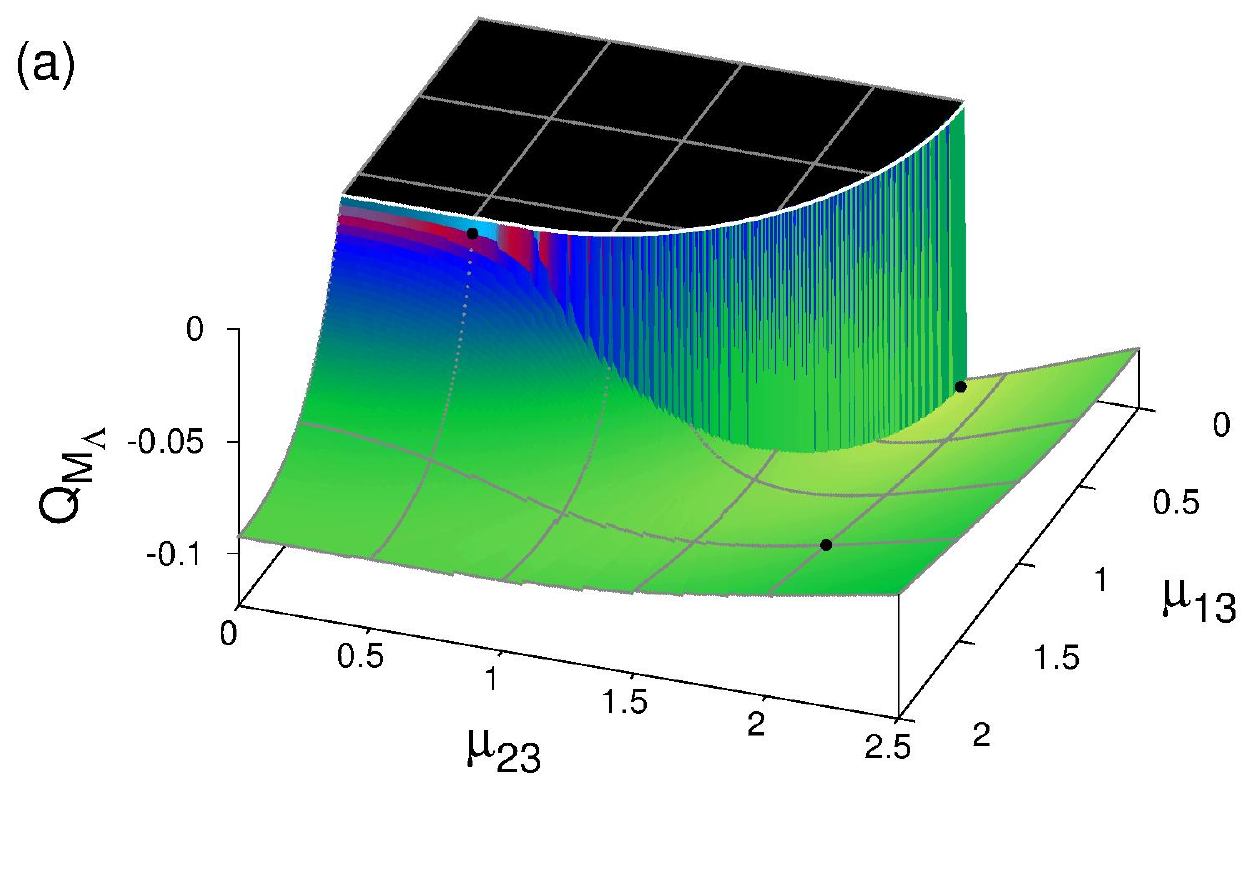}\
\includegraphics[width=0.48\linewidth]{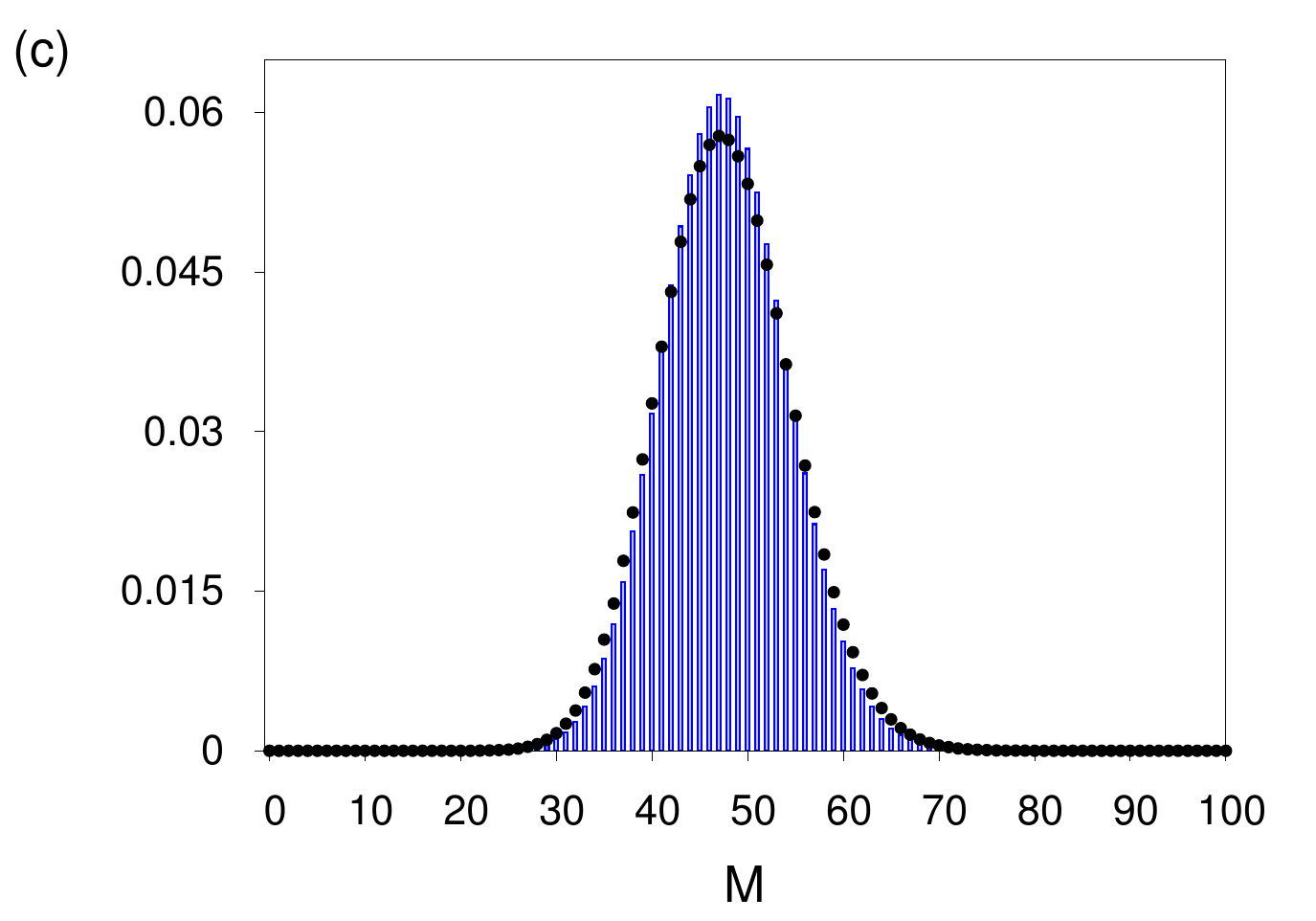}\\
\includegraphics[width=0.48\linewidth]{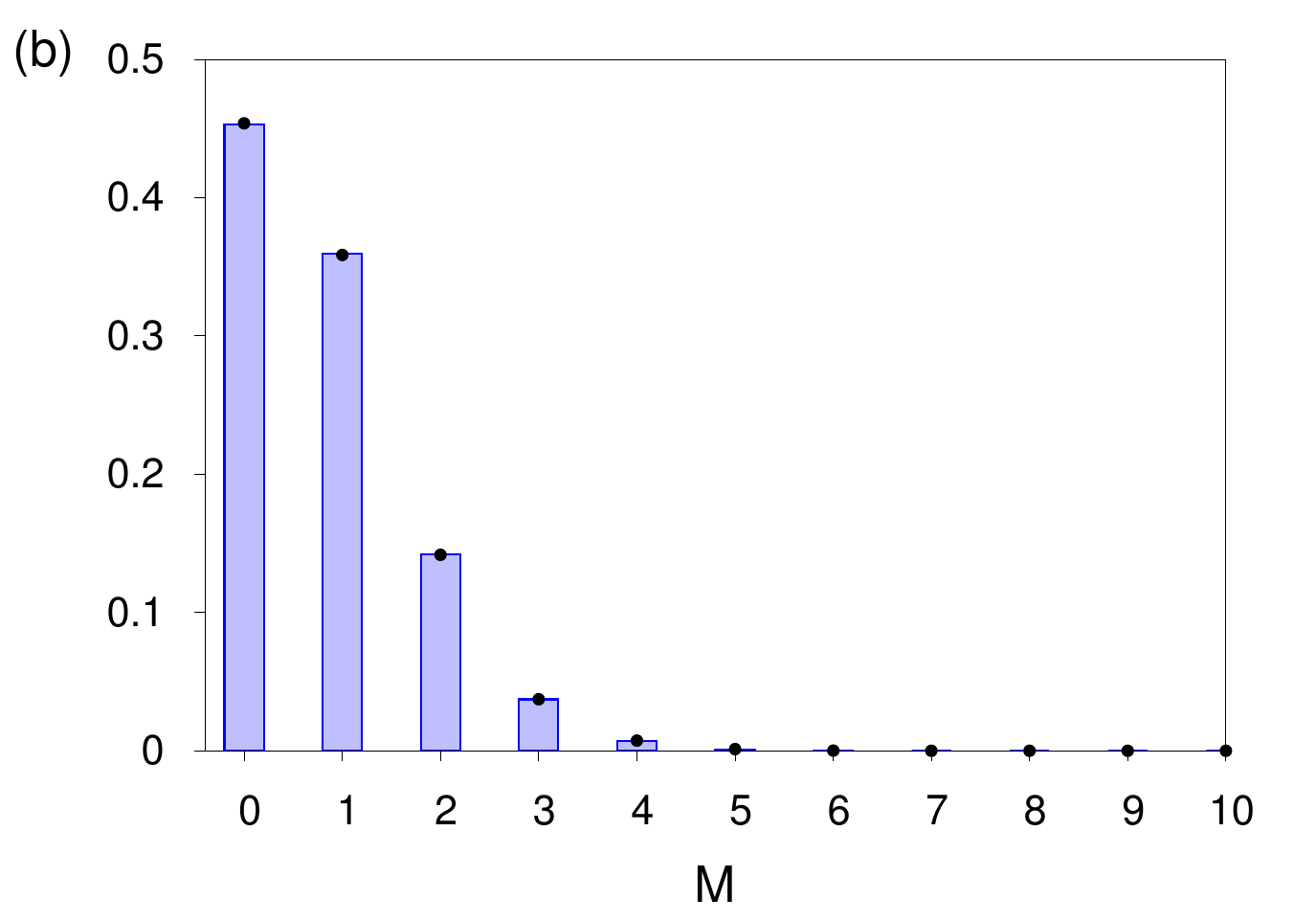}\
\includegraphics[width=0.48\linewidth]{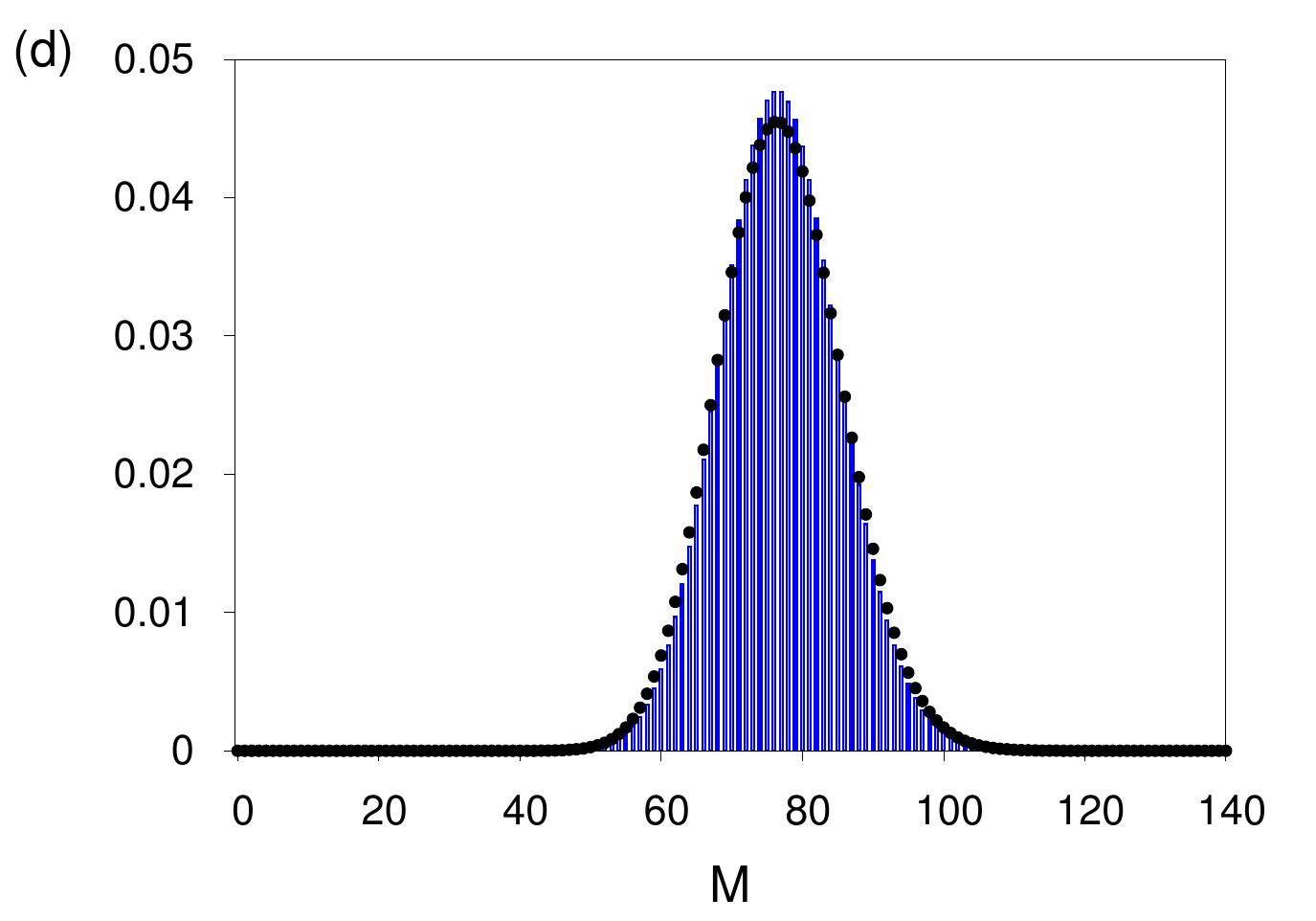}
\end{center}
\caption{(Color online.) (a) $Q_M$-Mandel parameter as a function of the control parameters, for atoms in the $\Lambda$ configuration in a non-resonant condition $\Delta_{31}=0.3$ and $\Delta_{32}=-0.2$. The separatrix is shown by a white line, and three points (dots) are shown where the corresponding ${\cal M}$ distribution of the ground state for $N_a=40$ atom has been calculated (solid bars) and compared with its corresponding Poissonian distribution (dots).  (b) ${\cal M}$ distribution for $\mu_{13} = 1.15,\ \mu_{23}=0.05 $, for which $M^c_\Lambda \approx 1.98\times 10^{-2}$ and $Q_M \approx -6.4\times10^{-3}$. (c) ${\cal M}$ distribution for $\mu_{13} = 0.05,\ \mu_{23}= 1.85$, for which $M^c_\Lambda \approx 1.19$ and $Q_M \approx -0.12$. (d) ${\cal M}$ distribution for $\mu_{13} = 1.5,\ \mu_{23}=2.0 $, for which $M^c_\Lambda \approx 1.92$ and $Q_M \approx -0.09$.}\label{f5}
\end{figure*}
%

%Figura 6
\begin{figure}
\begin{center}
\includegraphics[width=0.7\linewidth]{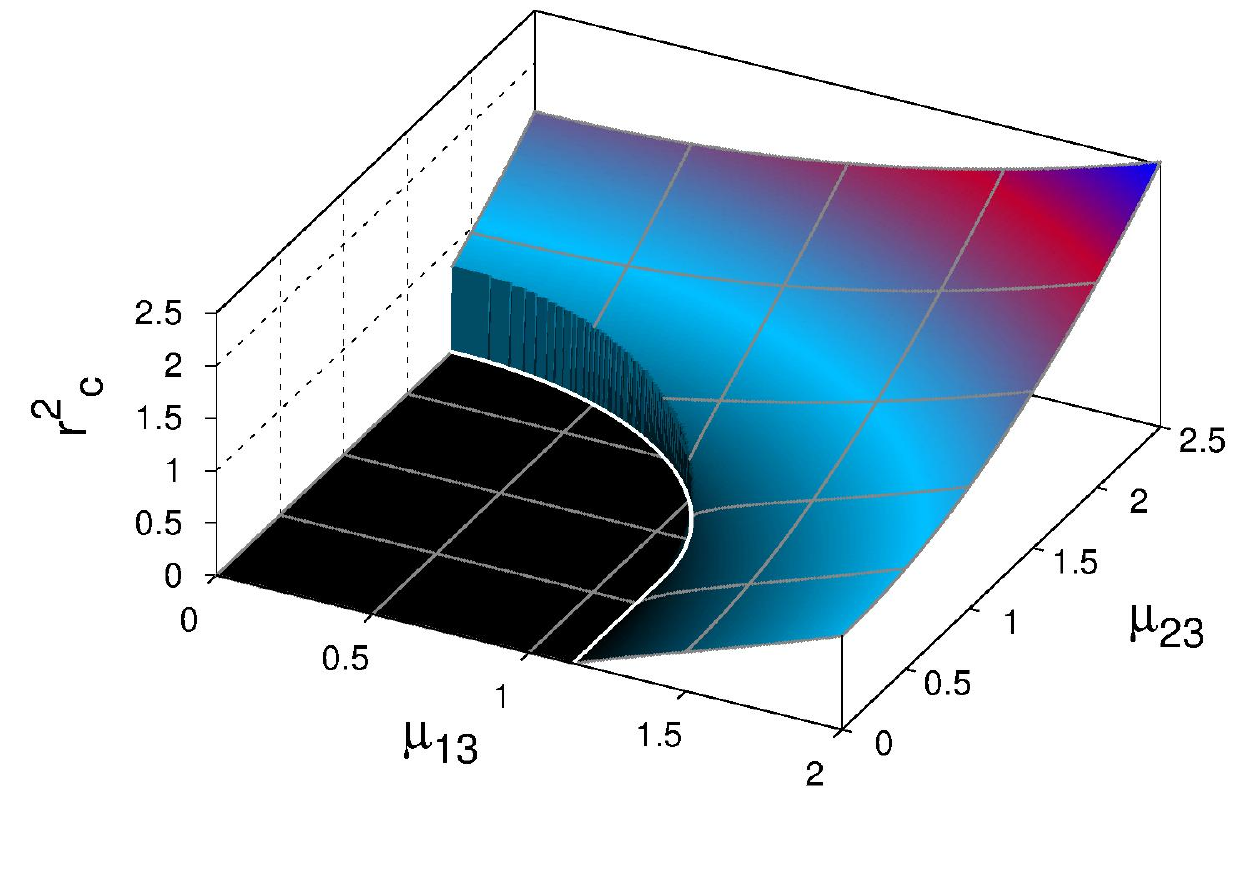}
\end{center}
\caption{(Color online.) Average value of the photon distribution in units of the total number of atoms $r^2_c = \rho^2_c/N_a$ [Cf. Eq. (\ref{eq.rho.minG})], for atoms in the $\Lambda$ configuration in the non-resonant case $\Delta_{31} = 0.3$ and $\Delta_{32}=-0.2$.}\label{f6}
\end{figure}

For atoms in the $\Lambda$ configuration it is required that the transitions from $\omega_1\longleftrightarrow\omega_2$ be negligible, and so we take $\mu_{12}=0$. The detuning for the corresponding values of the frequencies $\omega_2$ and $\omega_3$ are
%
%\begin{subequations}
\begin{eqnarray}
\omega_2= \Delta_{31} - \Delta_{32} + \omega_1,  \\
\omega_3 = \Delta_{31} + \omega_1 + \Omega.
\end{eqnarray}
%\end{subequations}
%
Because of the convention $\omega_1\leq\omega_2\leq\omega_3$ used in the labeling of the energy levels, the condition $\omega_1\approx\omega_2$ requires $\Delta_{31}-\Delta_{32}\approx 0$ with $\Delta_{31}\geq\Delta_{32}$.

First we consider the case of equal detuning, i.e., $\Delta_{31}=\Delta_{32}$. In this case, the critical points may be calculated analytically as functions of the control parameters. These are given by $\varrho_{2c}=\varrho_{3c} = 0$ in the normal regime, with ${\mu_{13}}^2+{\mu_{23}}^2\leq\Omega\,\omega_3$; while in the collective regime we have
%
%\begin{subequations}
\begin{eqnarray}\label{eq.L.criticos}
\varrho_{2c} = \frac{1}{\mu_{13}}\sqrt{\frac{\left(\mu_{13}^2+\mu_{23}^2\right)
\left(\mu_{13}^2+\mu_{23}^2-\Omega \, \omega_3\right)}{ \mu_{13}^2+\mu_{23}^2+\Omega \, \omega_3}}, \quad \\
\varrho_{3c} = \frac{\mu_{23}}{\mu_{13}},
\end{eqnarray}
%\end{subequations}
%
where states with ${\cal M}>0$ contribute to the ground state.

Substituting the critical points in the expression for the energy one finds that the minimum energy surface is given by $E_\Lambda^c=0$ for $\mu_{13}^2+\mu_{23}^2\leq\Omega \, \omega_3$ and
\begin{eqnarray}\label{eq.L.minE}
E^c_{\Lambda} = -\frac{1}{4\Omega}\frac{\left(\mu_{13}^2+\mu_{23}^2-\Omega \, \omega_3\right)^2}{ \mu_{13}^2+\mu_{23}^2},
\end{eqnarray}
in the collective region. Taking the first derivatives of the minimum energy surface and evaluating at the separatrix, one finds that only second-order transitions occur.

In the collective regime the $Q_M$-Mandel parameter reads
\begin{eqnarray}\label{eq.QM.L}
Q_M = - \frac{\Omega^2 \left(\mu_{13}^2 + \mu_{23}^2 - \Omega \, \omega_3\right)}{\left(\mu_{13}^2 + \mu_{23}^2 \right)  \left(\mu_{13}^2 + \mu_{23}^2 + \Omega \, \omega_3\right)}.
\end{eqnarray}
One can show that, independently of the detuning values, $Q_M=0$ in the normal regime (${\cal M}=0$), yielding Poissonian statistics, while in the collective regime (${\cal M}>0$) we have sub-Poissonian statistics, $Q_M<0$. Also we notice that $Q_M\to 0 $ when the control parameters go to infinity.   

We now consider atoms in the $\Lambda$ configuration with $\Delta_{31}\neq\Delta_{32}$, we chose $\Delta_{31} = 0.3$ and $\Delta_{32} =-0.2$. In this case the problem does not have an analytic solution and one needs to consider numerical solutions as for the $\Xi$ configuration.

Figure \ref{f4} shows the first derivatives of the semi-classical energy surface for the ground state. 
These present discontinuities along the separatrix  where $|\mu_{23}|>\sqrt{\Omega \, \omega_{21}}$  indicating first-order transitions. In the region where $|\mu_{23}|<\sqrt{\Omega \, \omega_{21}}$  with  $\mu_{13} = \sqrt{\Omega \, \omega_{31}}$ the derivatives are continuous, and second-order transitions occur. The corresponding $Q_M$-Mandel parameter  and the $M$-distribution of the coherent state for three values with $N_a=40$ atoms is shown in Fig.~\ref{f5}. Here we compare the sub-Poissonian distribution of the state with its corresponding Poissonian distribution (dots). Finally the photon number distribution is shown in Fig.~\ref{f6}. One should compare the behavior of these quantities Figs. \ref{f4}, \ref{f5} and \ref{f6} with the corresponding ones for the $\Xi$ configuration, Figs. \ref{f1}, \ref{f2} and \ref{f3}, respectively. Notice that the behavior is very similar, i.e., for atoms in $\Lambda$ configuration with unequal detuning, the physical quantities and properties (order of the transitions) resemble those of the atoms in $\Xi$ configuration: they are both qualitatively equivalent.

\subsubsection{${\rm V}$ configuration}\label{NR.V}
% 

%Figura 7
\begin{figure}
\begin{center}
\includegraphics[width=0.48\linewidth]{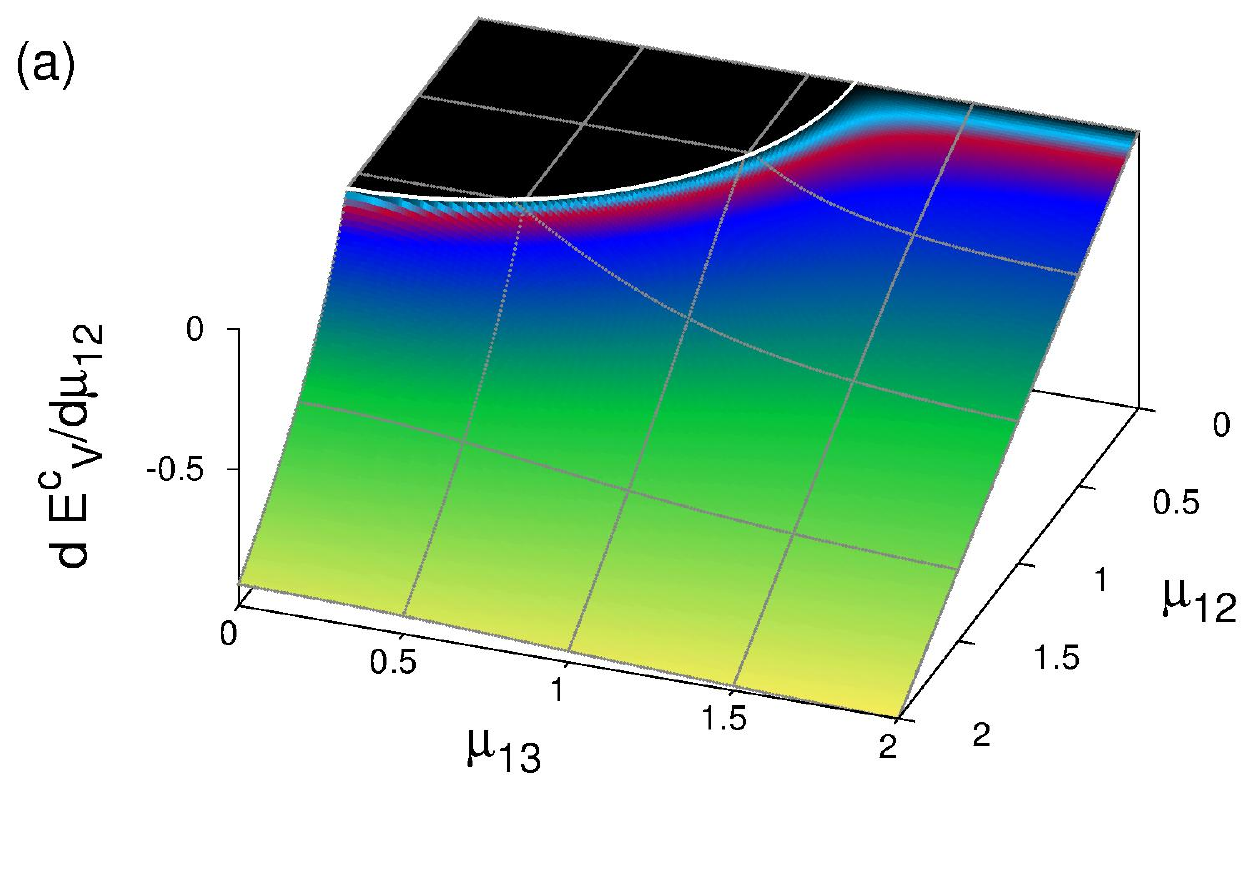}\,
\includegraphics[width=0.48\linewidth]{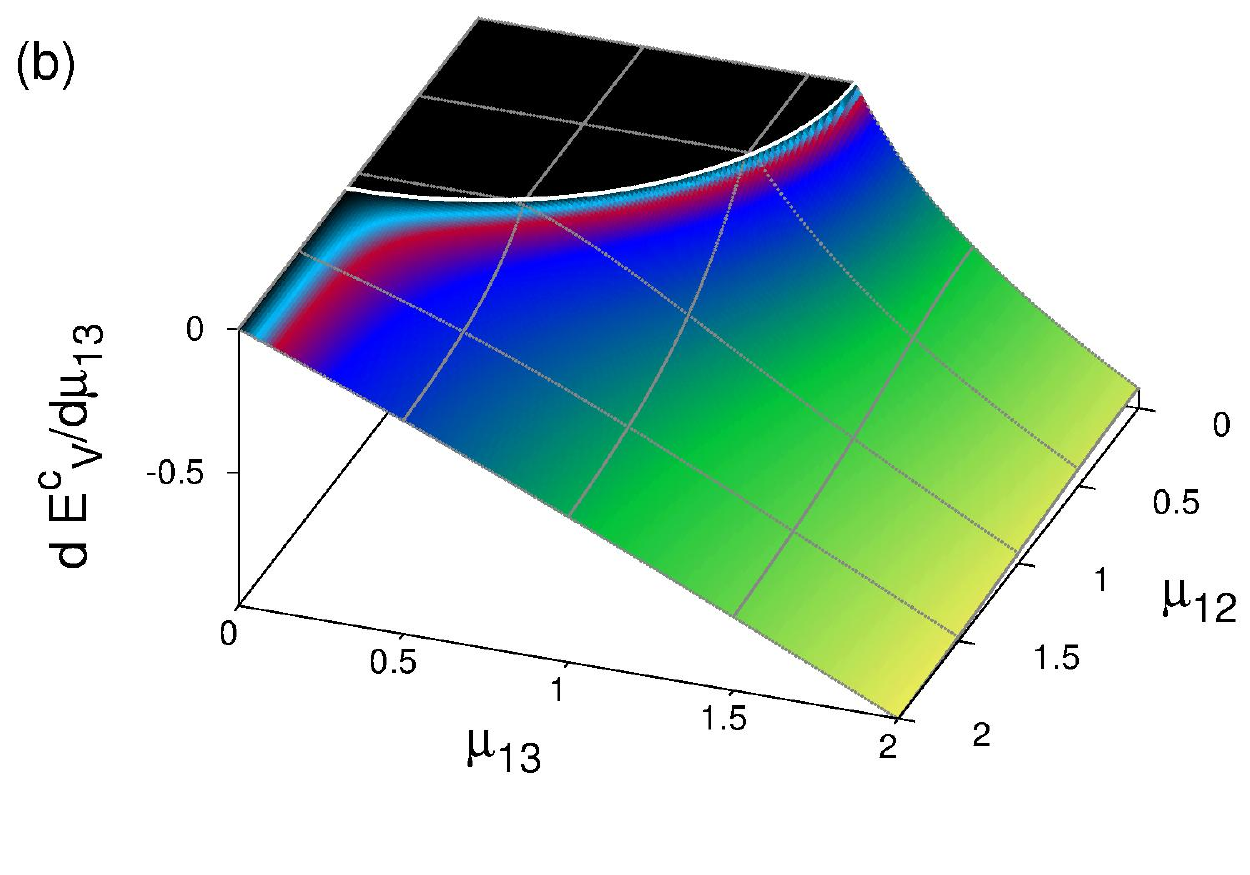}
\end{center}
\caption{(Color online.) First derivatives of the ground state energy with respect to its control parameters, for atoms in the $V$ configuration with a $\Delta_{21}=0.2$ and $\Delta_{31}=0.3$ detuning. (a) derivative with respect to $\mu_{12}$ and (b) derivative with respect to $\mu_{13}$. }\label{f7}
\end{figure}
% Figura 8
\begin{figure*}
\begin{center}
\includegraphics[width=0.48\linewidth]{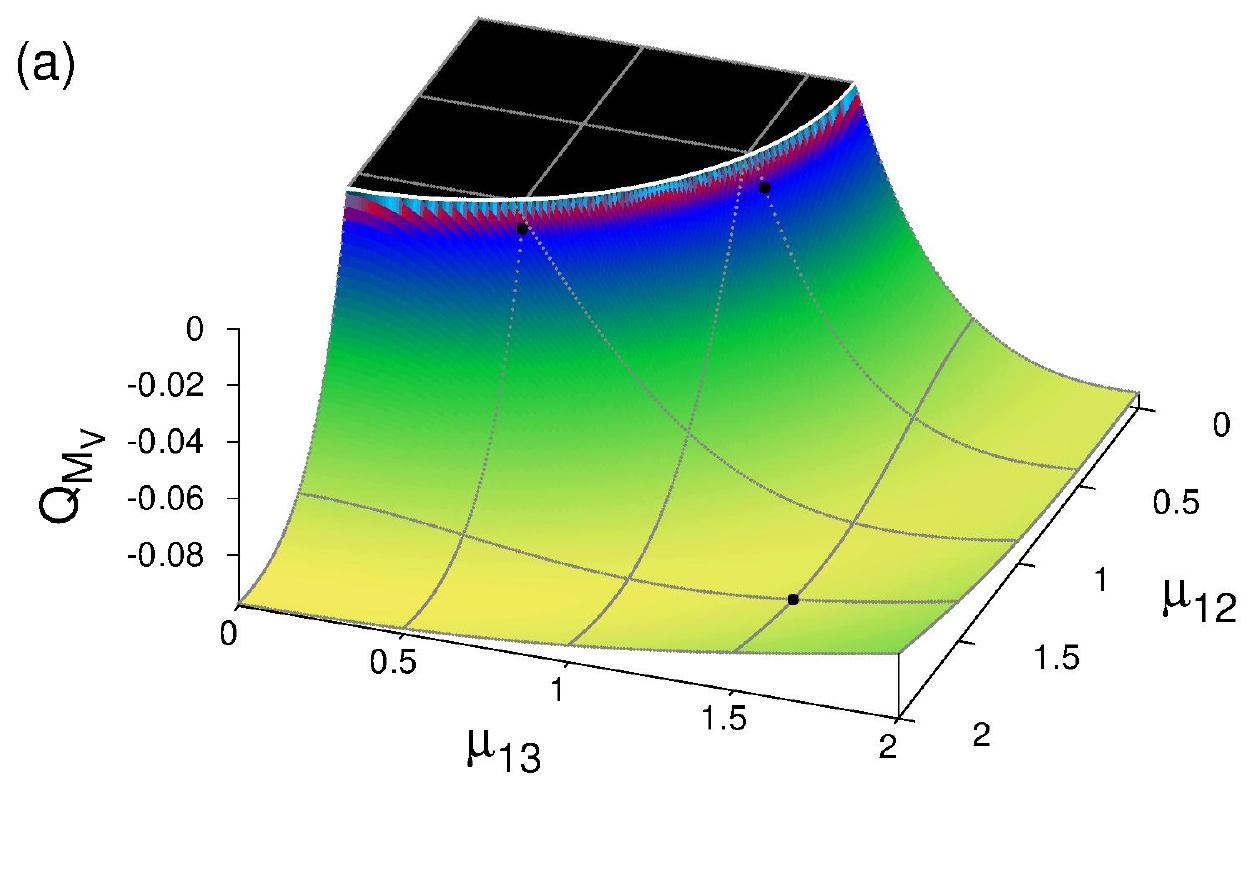}\
\includegraphics[width=0.48\linewidth]{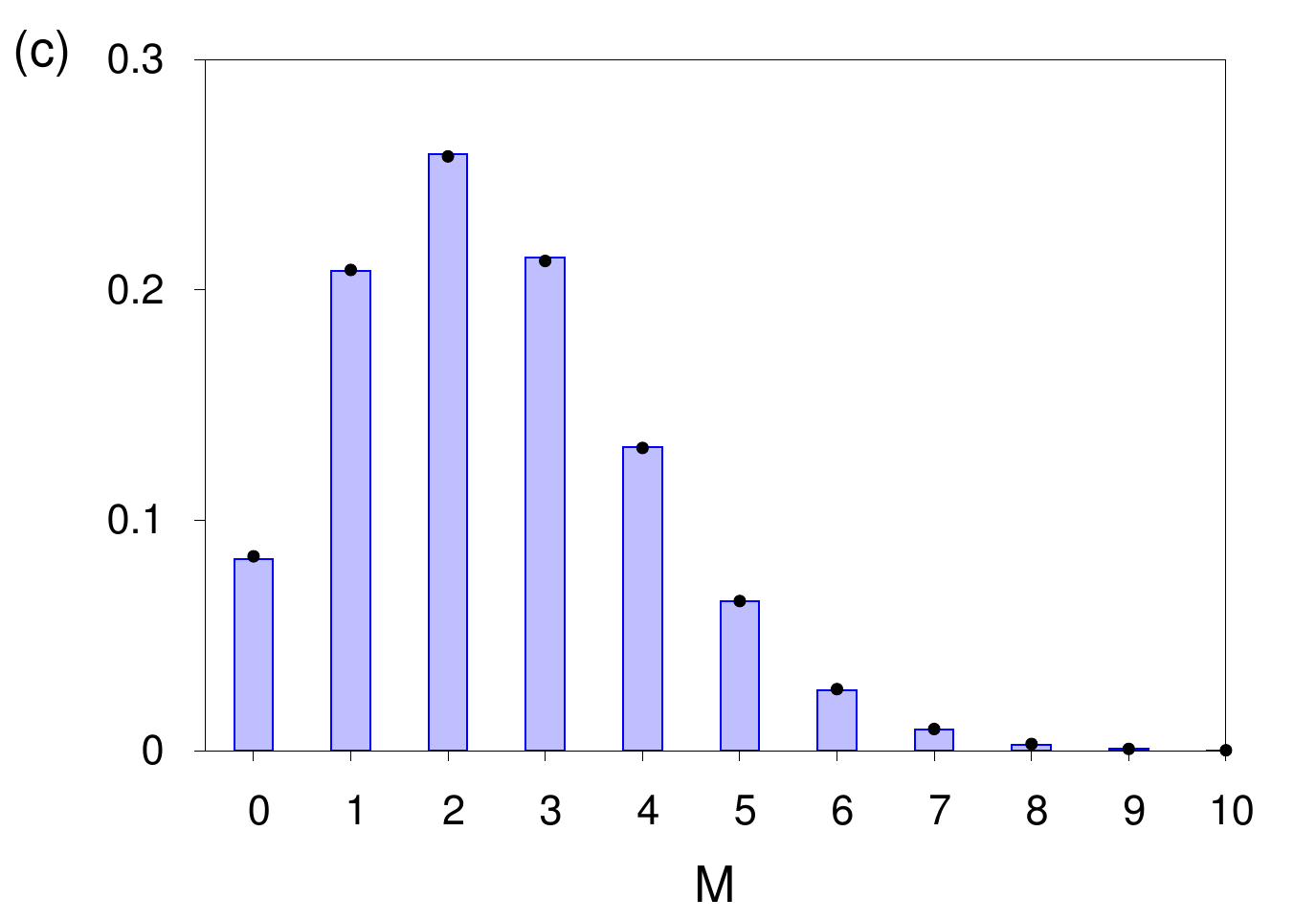}\\
\includegraphics[width=0.48\linewidth]{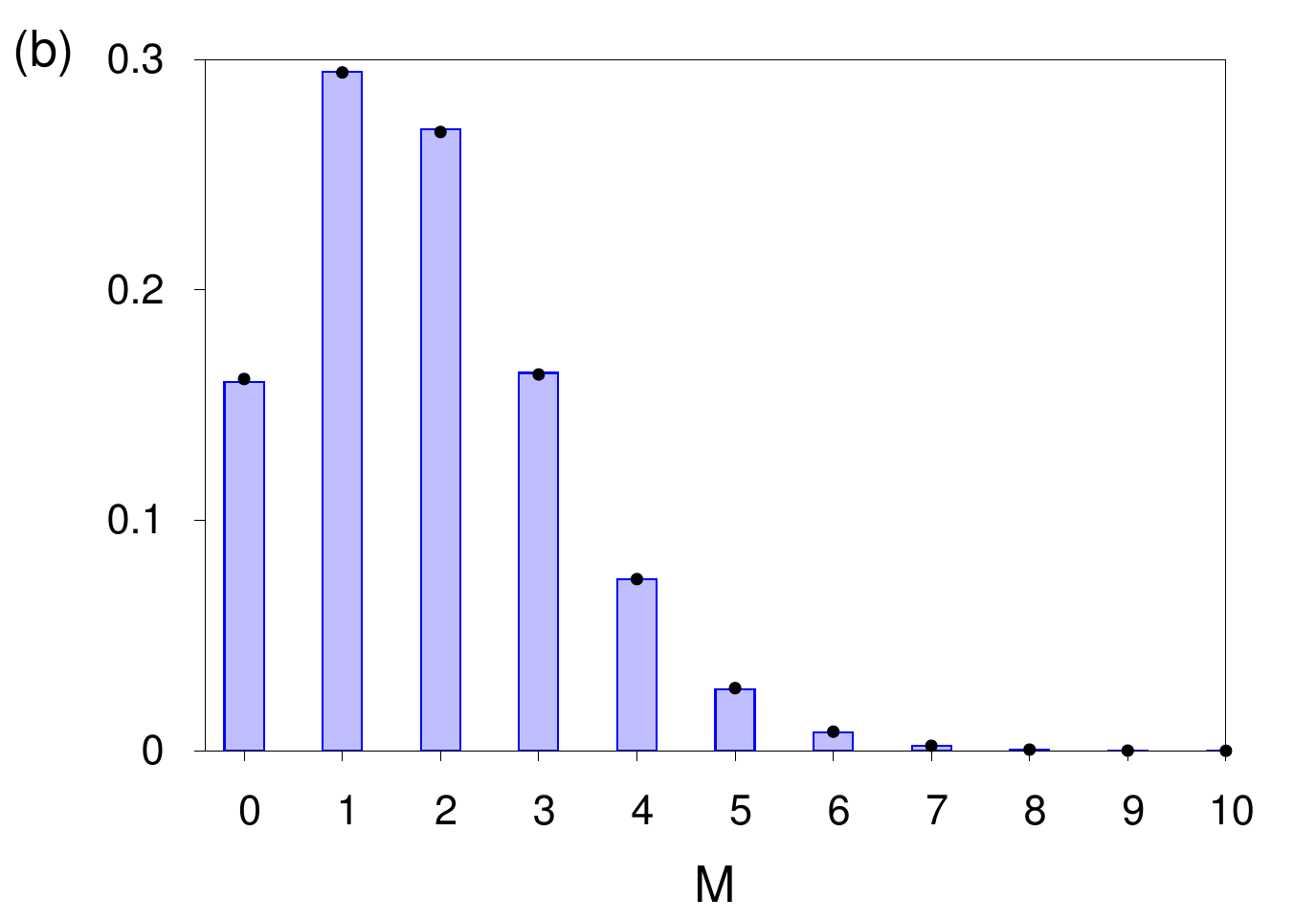}\
\includegraphics[width=0.48\linewidth]{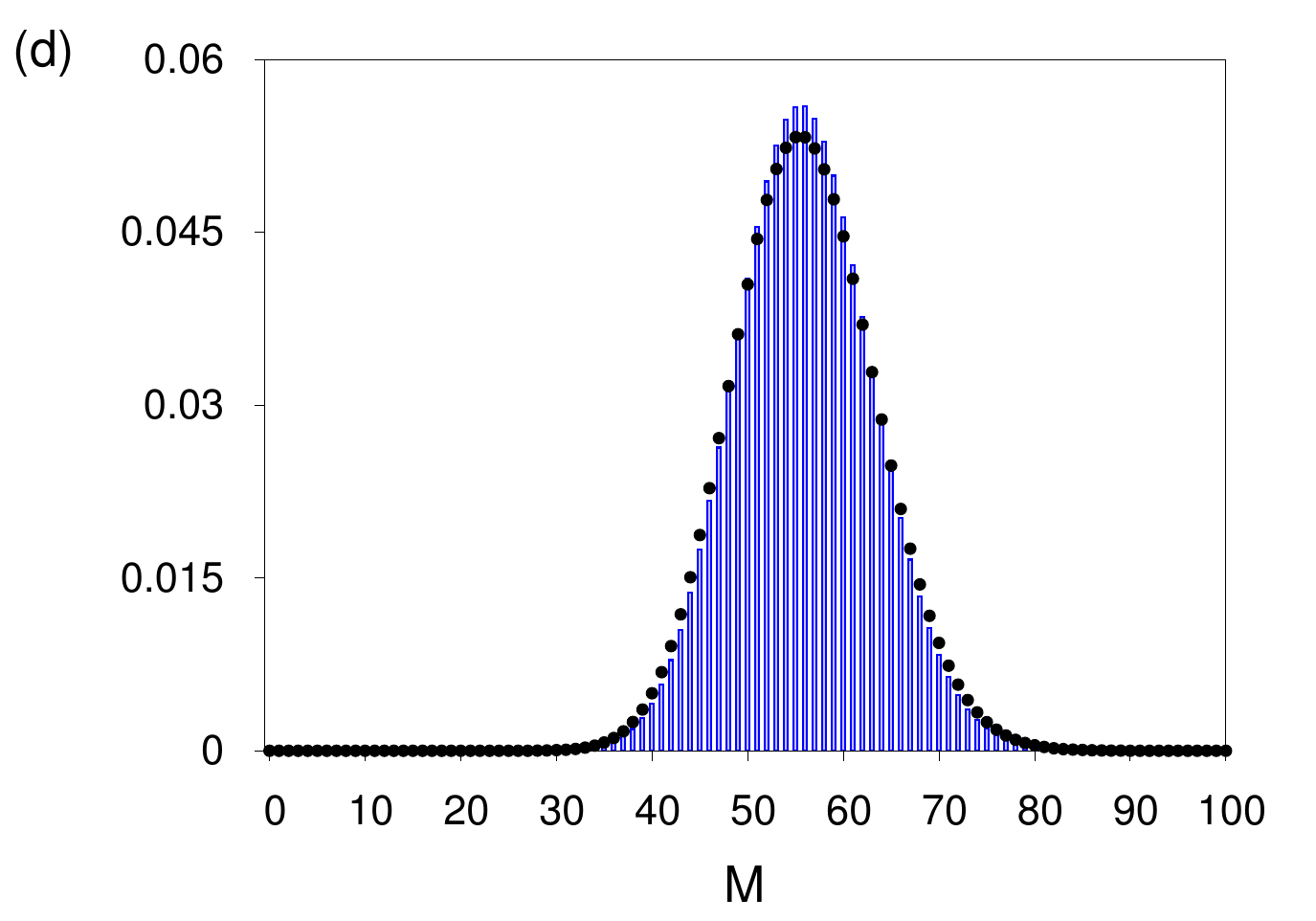}
\end{center}
\caption{(Color online.) (a) $Q_M$-Mandel parameter as a function of the control parameters, for atoms in the $V$ configuration in the non-resonant case $\Delta_{21}=0.2$ and $\Delta_{31}=0.3$. The separatix is shown by a white line, and three points (dots) are displayed where the corresponding ${\cal M}$ distribution of the ground state for $N_a=40$ atom has been calculated (solid bars) and compared with its corresponding Poissonian distribution (dots).  (b) ${\cal M}$ distribution for $\mu_{12}= 1.01,\ \mu_{13}=0.5 $, for which $M^c_V \approx 4.56\times 10^{-2}$ and $Q_M \approx-9.06 \times 10^{-3} $. (c) ${\cal M}$ distribution for $\mu_{12} = 0.5,\ \mu_{13}= 1.05$, for which $M^c_V \approx 6.18\times 10^{-2}$ and $Q_M \approx -1.15\times 10^{-2}$. (d) ${\cal M}$ distribution for $\mu_{12} = 1.5,\ \mu_{13}=1.5 $, for which $M^c_V \approx 1.4$ and $Q_M \approx -9.33\times10^{-2}$.}\label{f8}
\end{figure*}
%

%Figura 9
\begin{figure}
\begin{center}
\includegraphics[width=0.7\linewidth]{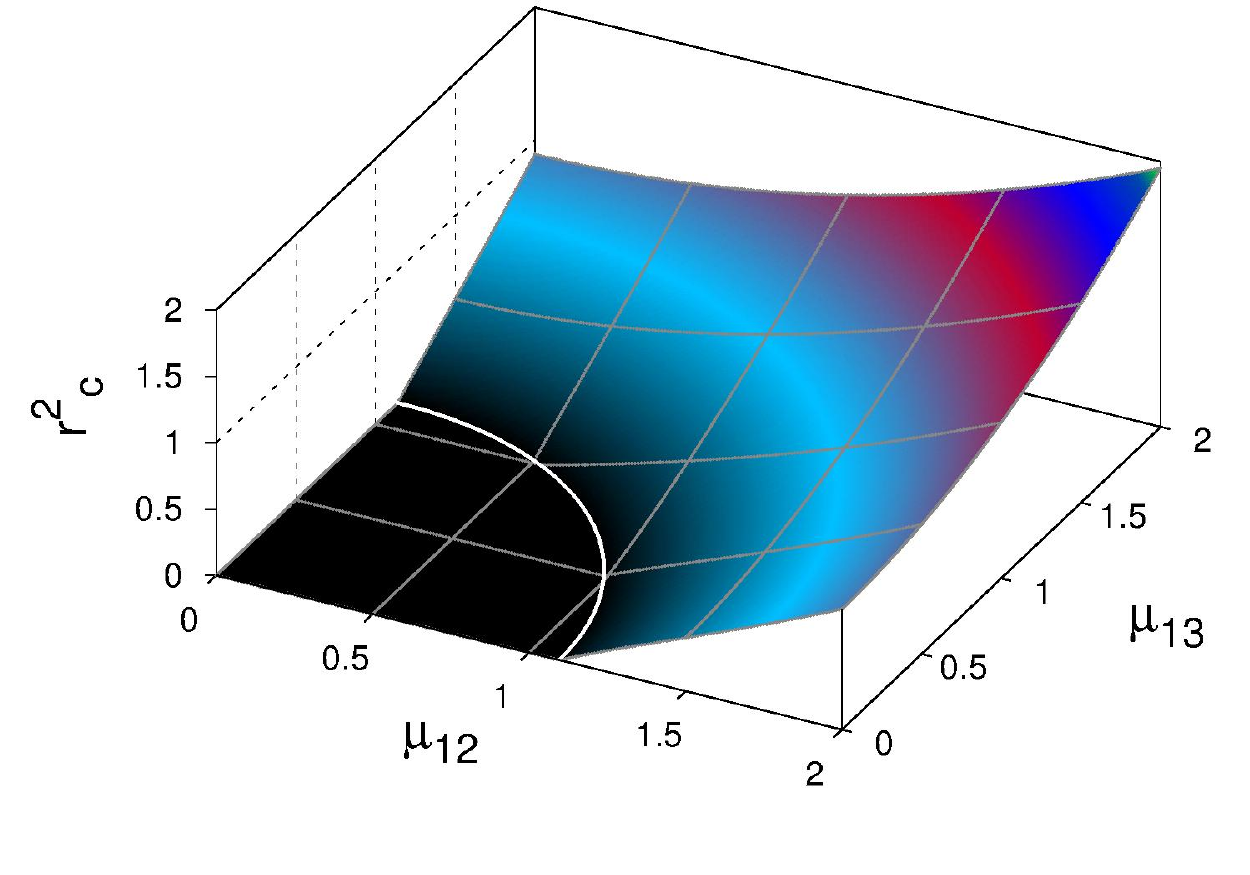}
\end{center}
\caption{(Color online.) Average value of the photon number distribution in units of the total number of atoms $r^2_c = \rho^2_c/N_a$ [see Eq. (\ref{eq.rho.minG})], for atoms in the $V$ configuration considering the non-resonant case $\Delta_{21}=0.2$ and $\Delta_{31}=0.3$.}\label{f9}
\end{figure}
%

% Figura 10
\begin{figure*}
\begin{center}
\includegraphics[width=0.48\linewidth]{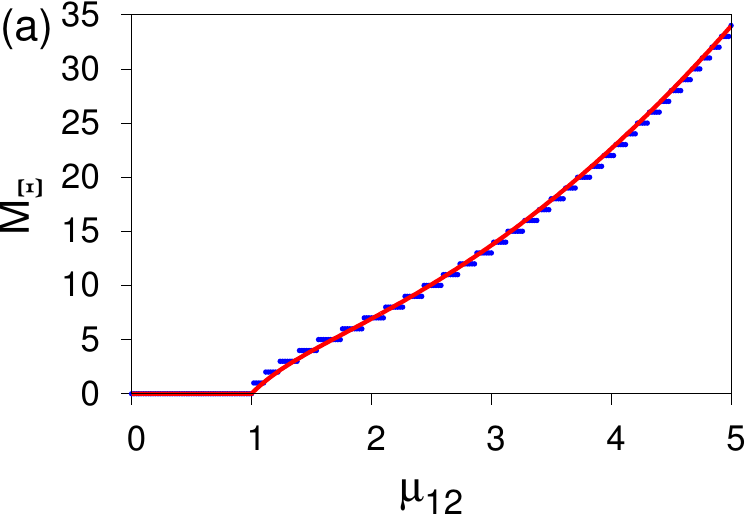}\,
\includegraphics[width=0.48\linewidth]{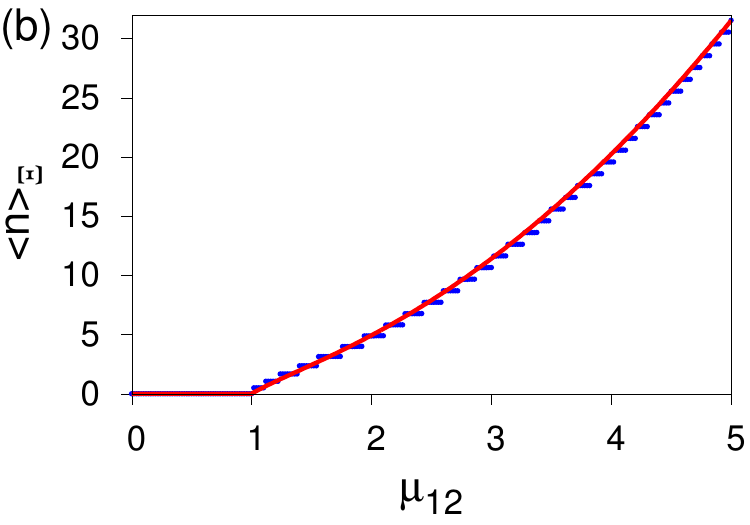}\\[3mm]
\includegraphics[width=0.48\linewidth]{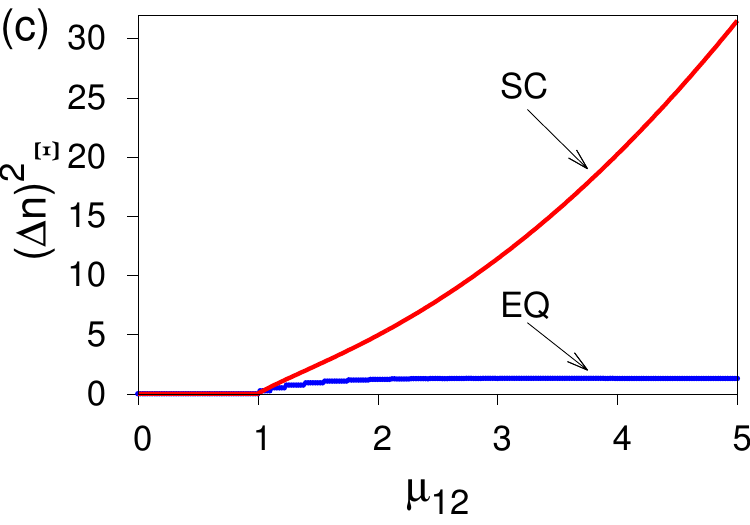}
\end{center}
\caption{(Color online.) Properties of atoms in the $\Xi$ configuration in a double resonance condition, with fixed value $\mu_{23}=0.5$ and $N_a=5$, in the semi-classical (SC, continuous line) and exact quantum calculation (EQ, dots) are compared as functions of the control parameter $\mu_{12}$. (a) Expectation value of the total number of excitations $\langle \bm{M}\rangle={\cal M}_\Xi$, (b) expectation value of the number of photons $\langle \bm{n}\rangle$, and (c) the corresponding fluctuations of the number of photons $\left(\Delta n\right)^2_\Xi$. }\label{f10}
\end{figure*}

A system of atoms in the ${\rm V}$ configuration requires $\mu_{23}=0$, since transitions between the levels $\omega_3$ and $\omega_2$ are negligible. In this case, the detuning parameters are $\Delta_{21}$ and $\Delta_{31}$  are given by     
%
%\begin{subequations}
\begin{eqnarray}
\omega_2 = \Delta_{21}+\omega_1 +\Omega,\\
\omega_3 = \Delta_{31}+\omega_1 + \Omega.
\end{eqnarray}
%\end{subequations}
%
Notice that the condition $\omega_2\approx\omega_3$ on $\omega_1 \leq\nobreak \omega_2\leq\omega_3$ reads, in terms of the detuning, as $\Delta_{21}\approx\Delta_{31}$ but satisfying $\Delta_{21}\leq\Delta_{31}$.

In a similar form to the atoms in the $\Lambda$ configuration, when the detuning parameters are equal, $\Delta_{21}=\Delta_{31}$, the problem has analytic solution. The critical points are $\varrho_{2c}=\varrho_{3c}=\nobreak0$ for the normal regime implying an energy surface for the ground state equal to zero.  For the collective regime, $\varrho_{2c}$ and $\varrho_{3c}$ take the values
%
%\begin{subequations}
\begin{eqnarray}\label{eq.V.criticos}
\varrho_{2c} = \mu_{13} \sqrt{\frac{\mu_{12}^2+ \mu_{13}^2-\Omega \, \omega_3}{\left(\mu_{12}^2+ \mu_{13}^2\right) \left(\mu_{12}^2+\mu_{13}^2+\Omega \, \omega_3\right)}}, \quad\\
\varrho_{3c} =\mu_{12}\sqrt{\frac{\mu_{12}^2+ \mu_{13}^2 - \Omega \, \omega_3}{\left(\mu_{12}^2+\mu_{13}^2\right) \left(\mu_{12}^2+\mu_{13}^2 + \Omega \, \omega_3\right)}} \, . \quad
\end{eqnarray}
%\end{subequations}
%
Substituting these into the expression for the energy Eq. (\ref{eq.E.G3}), one finds 
\begin{eqnarray}\label{eq.V.minE}
E^c_{V} = -\frac{1}{4\Omega} \frac{\left(\mu_{12}^2 + \mu_{13}^2 - \Omega \, \omega_3\right)^2}{\mu_{12}^2 + \mu_{13}^2} \, .
\end{eqnarray}
This is similar as for atoms in the $\Lambda$ configuration, in fact, the expression is equal by just replacing $\mu_{23}\to\mu_{12}$ in Eq. (\ref{eq.L.minE}). A similar situation occurs for the $Q_M$-Mandel parameter, which is given by    
\begin{eqnarray}
Q_M = - \frac{\Omega^2\left(\mu_{12}^2 + \mu_{13}^2 - \Omega \, \omega_3\right)}{\left(\mu_{12}^2 + \mu_{13}^2 \right)\left(\mu_{12}^2 + \mu_{13}^2 + \Omega \, \omega_3\right)} \, .
\end{eqnarray}
Hence, atoms in both configurations $V$ and $\Lambda$ have similar properties under equal detuning considerations.

By considering the case of unequal detuning $\Delta_{21}\neq\Delta_{31}$, we choose to analyze the case $\Delta_{21} = 0.2$ and $\Delta_{31}=0.3$. Fig. \ref{f7} shows the first derivatives of the energy surface for the ground state as a function of the control parameters $\mu_{12},\ \mu_{13}$.  In both cases the first derivative is continuous, and so second-order transitions are present. 

Similarly the $Q_M$-Mandel parameter is continuous [Fig.~\ref{f8}(a)] in a vicinity of the separatrix (white line). The corresponding ${\cal M}$ distribution of the coherent state with $N_a=40$ for three different points are shown in Figs.~\ref{f8}(b), (c) and (d) (bars), and these are compared with their respective Poissonian distribution (dots). One can observe that the ${\cal M}$ distribution is very close to the corresponding Poissonian one, and this is due to the fact that $Q_M\sim 10^{-2}$ is close to zero for any considered value.

Finally, Fig.~\ref{f9} shows the corresponding photon number distribution in units of the total number of atoms $r_c^2 = \rho_c^2/N_a$. This quantity is a continuous smooth function around the separatrix, since this configuration presents only second-order transitions. 

The same results are obtained for various values of the detuning parameters.

% Figura 11
\begin{figure*}
\begin{center}
\includegraphics[width=0.48\linewidth]{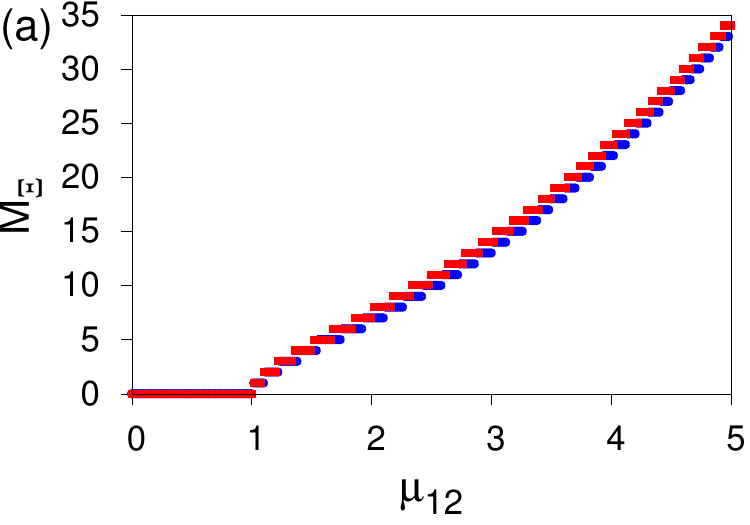}\,
\includegraphics[width=0.48\linewidth]{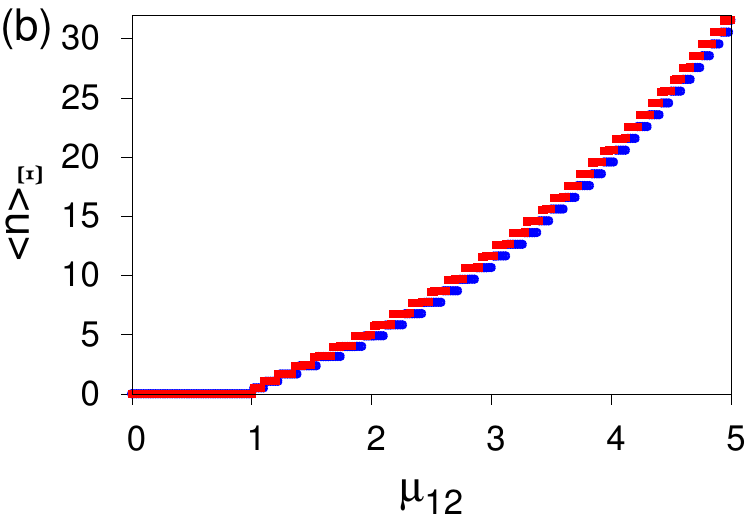}\\[3mm]
\includegraphics[width=0.48\linewidth]{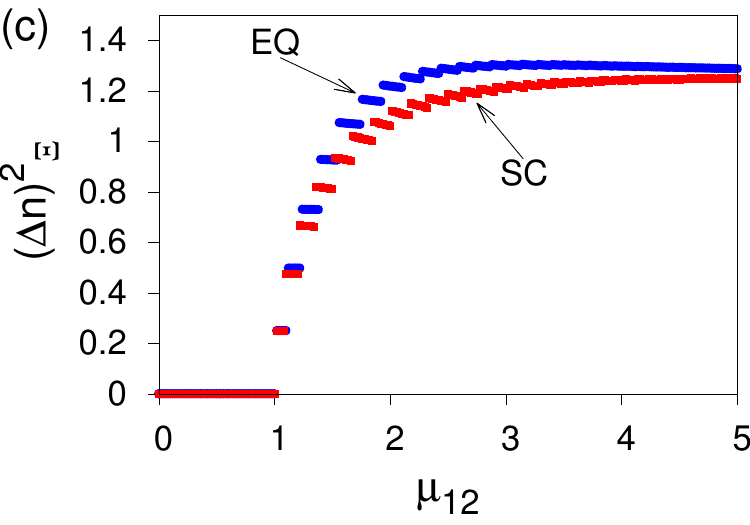}
\end{center}
\caption{(Color online.) Properties of atoms in the $\Xi$ configuration in a double resonance condition, with fixed value $\mu_{23}=0.5$ and $N_a=5$, in the projected semi-classical state (SC, lighter dots) and the exact quantum calculation (EQ, darker dots) are compared as functions of the control parameter $\mu_{12}$. (a) Expectation value of the total number of excitations $\langle \bm{M}\rangle={\cal M}_\Xi$, (b) expectation value of the number of photons $\langle \bm{n}\rangle$, and (c) the fluctuations of the number of photons $\left(\Delta n\right)^2_\Xi$.}\label{f11}
\end{figure*}

\section{Comparison with the quantum solution}\label{QNS}

The exact numerical calculation of the ground state energy may be evaluated using the uncoupled basis given by the direct product between the field $|n\rangle$ and matter states Eq. (\ref{eq.G.state}). Since we have chosen $h_1=N_a$ and $h_2=h_3=0$ one may simplify the Gelfand-Tsetlin notation as 
\begin{eqnarray}\label{eq.uncoupled.basis}
|n q_1 r\rangle \equiv |n\rangle \otimes\left|\begin{array}{c c c} q_1 & & 0 \\ & r & \end{array}\right\rangle \, ,
\end{eqnarray}
$q_2$ is zero because it must satisfy $h_2\geq q_2\geq h_3$. The corresponding matrix elements of the operators $\bm{A}_{ij}$ (for this particular basis) are given in the appendix \ref{ap.Aij}, which can be used to calculate the matrix elements of the Hamiltonian, Eq. (\ref{eq.H.3level}), and to evaluate numerically its eigenvalues. 

For each particular atomic configuration ($\Xi,\ \Lambda$ or ${\rm V}$) there is an additional constant of motion $\bm{M}$, namely total number of excitations (\ref{eq.M.Xi}-\ref{eq.M.V}). Taking a particular configuration, the Hamiltonian has a matrix representation as a {\em block diagonal matrix}, where the dimension of each matrix of the diagonal depends of $M^q$ and $N_a$. For large values of $M^q$, however, the dimension depends only of $N_a$ and is given by
\begin{equation*}
\frac{N_a(N_a +1)}{2} + N_a +1 \, ;
\end{equation*}
this occurs for $M^q_\Xi\geq 2 \, N_a$ ($\Xi$ configuration), $M^q_\Lambda\geq  N_a$ ($\Lambda$ configuration) and $M^q_V\geq N_a$ (${\rm V}$ configuration), relationships provided by the condition $n\geq0$ in Eq. (\ref{eq.M.Xi}-\ref{eq.M.V}). For $M^q_\Xi< 2N_a, \ M^q_\Lambda<N_a$ or $M^q_V<N_a$ we could not find a simple relationship for the dimension of matrix.

To find the quantum ground energy and its corresponding eigenstate, we proceed as follows. For each configuration of the atom, we take a value of $M^q$, and for fixed parameters $\Omega,\ \omega_1,\ \omega_2$ and $\omega_3$ the eigenvalues and their corresponding eigenstates are evaluated numerically as functions of the control parameters $\mu_{ij}$.  This gives us the ground state energy for each corresponding total number of excitations. 

It is worth mentioning that, for a fixed region of values of the interaction intensity, one may estimate the maximum value of $M^q$ that is required to find the minimum energy; this value is provided by semi-classical calculation.      

In order to see how well the semi-classical results approximate the corresponding exact quantum ones, we consider atoms in the $\Xi$ configuration in a double resonance for $N_a=5$ atoms. Notice that the quantum calculation of the ground state depends on the number of atoms $N_a$ considered, and this is in contrast with the semi-classical one where this quantity plays the role of an extensive variable. Let us focus on the expectation values of the total number of excitations $\langle  \bm{M}\rangle$, number of photons $\langle \bm{n}\rangle$ and its fluctuations $\left(\Delta n\right)^2=\langle \bm{n}^2\rangle-\langle \bm{n}\rangle^2$.

% Figura 12
\begin{figure*}
\begin{center}
\includegraphics[width=0.48\linewidth]{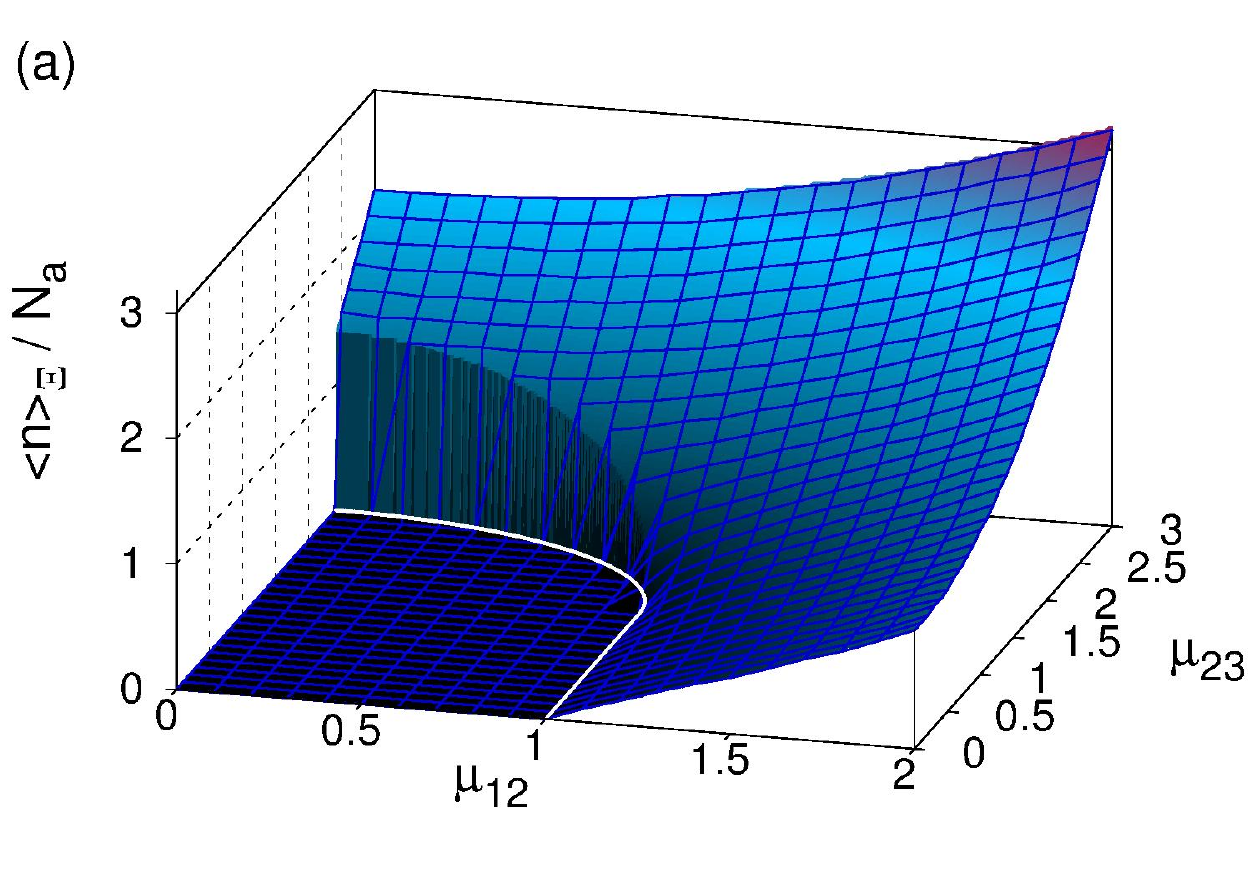} \,
\includegraphics[width=0.48\linewidth]{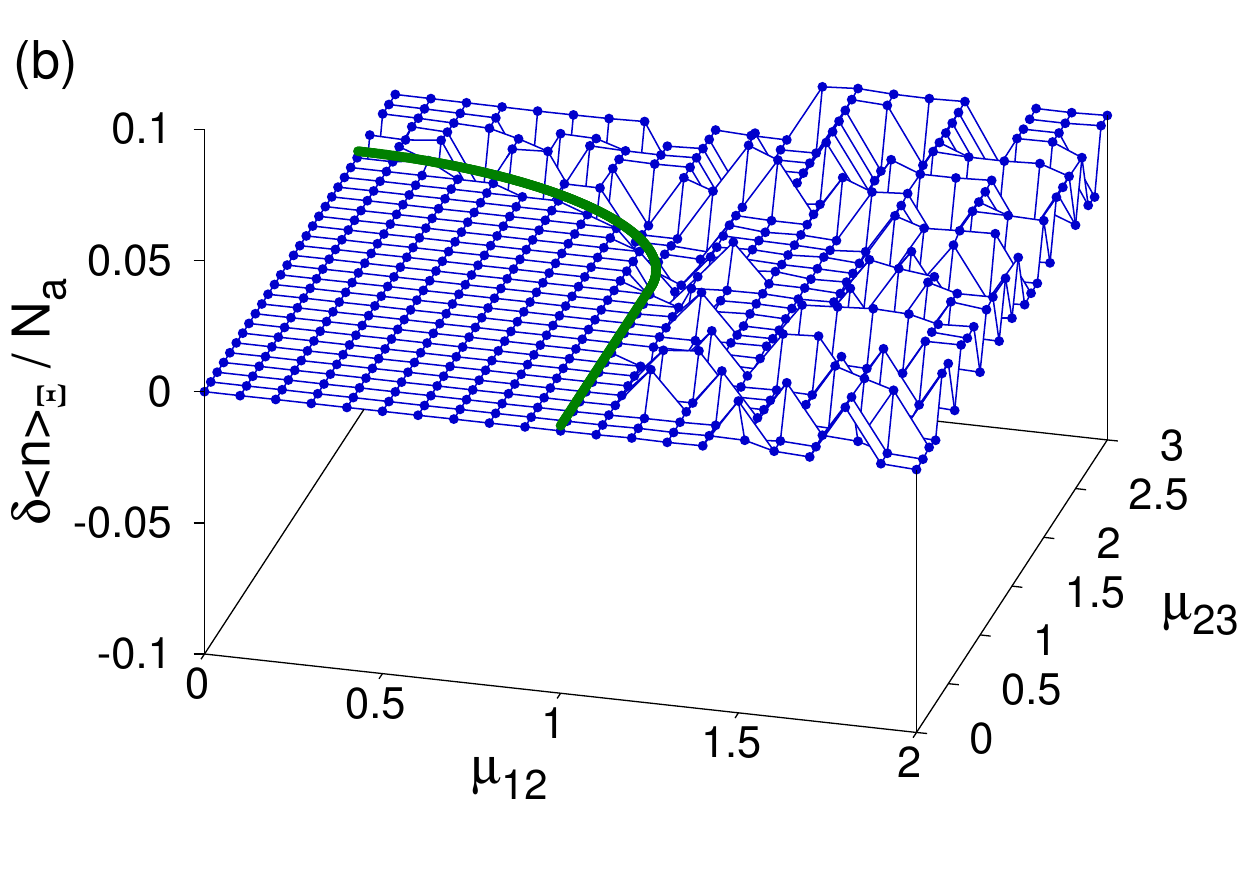} 
\end{center}
\caption{(Color online.) Exact and projected solutions compared for atoms in the $\Xi$ configuration in double resonance $\Delta_{21}=\Delta_{32}=0$, considering $N_a=40$ atoms. (a) expectation value of the number of photons, showing no visual differences, and (b)  the corresponding fluctuations. In both cases, the quantities were normalized to the number of atoms $N_a$.}\label{f12}
\end{figure*}

Fig.~\ref{f10} shows, respectively, the expectation values of the total number of excitations [Fig.~\ref{f10}(a)], number of photon [Fig.~\ref{f10}(b)] and photon fluctuations [Fig.~\ref{f10}(c)] as a function of the intensity $\mu_{12}$ for a fixed value $\mu_{23}=0.5$. In all cases, the semi-classical calculation is represented by a continuous line while the corresponding exact quantum calculation by dots. One may observe that in the case of the expectation values both calculations are in very good agreement [Figs.~\ref{f10} (a) and (b)].  The fluctuation in the number of photons however fails to render the quantum results [Fig.~\ref{f10}(c)], except in the normal regime where $\langle  \bm{n}\rangle = 0$ in both cases. This difference is due to the fact that in the semi-classical ground state a coherent state for the photon contribution is considered, and hence, the fluctuations are equal to its expectation value, $\left(\Delta n\right)^2 = \langle  \bm{n}\rangle$, in other words this possesses a Poissonian distribution. However, the photon distribution of the exact ground state does not have this property, because the total number of excitations is fixed for this state. 

The above comparison suggests that we should consider an additional correction to our semi-classical test state. 

\section{Projected variational state}\label{quantum.proj}

The matter unnormalized ${\rm U}(3)$ coherent state for the totally symmetric representation, i.e., $h_2=h_3=0$, can be written as 
\begin{equation}
|h_1, \, \vec{\gamma}\} = \sum^\infty_{n,m=0} \frac{\gamma_2{}^n}{n!} \,  \frac{\gamma_3{}^m}{m!} \, (\bm{A}_{21})^m \, (\bm{A}_{31})^n \vert h_1,0,0\rangle_F 
\end{equation}
because $\bm{A}_{32} \vert h_1,0,0\rangle_F =0$ and where $\vert h_1,0,0\rangle_F$ denotes the Gelfand-Tsetlin highest weight state (HWS).  In this case, one can represent the ${\rm U}(3)$ generators as follows: $\bm{A}_{31} = \bm{b}_3^\dagger \, \bm{b}_1$ and $\bm{A}_{21} = \bm{b}_2^\dagger \, \bm{b}_1$. Then the HWS can be written as
\begin{equation}\label{hws}
\vert h_1,0,0\rangle_F = \frac{1}{\sqrt {h_1!}} (\bm{b}_1^{\dagger})^{h_1} \vert 0,0,0 \rangle_F \, ,
\end{equation}
where we are using the Fock vacuum state $\vert 0,0,0 \rangle_F$ defined by $\bm{b}_k \vert 0,0,0 \rangle_F =0$ with $k=1,2,3$. The action of $\bm{A}_{31}$, and $\bm{A}_{21}$ on (\ref{hws}) is straightforward and results in 
%\begin{widetext}

\begin{equation*}
|h_1, \, \vec{\gamma}\} = \sum^{h_1}_{n=0} \, \sum^{h_1-n}_{m=0} {\textstyle\sqrt{\frac{h_1 !}{(h_1-n-m)! \, n! \, m!}}} \ 
 \gamma_2{}^m \,  \gamma_3{}^n\vert h_1-n-m,n,m \rangle_F \, .
\end{equation*}
Therefore, the semi-classical variational state constructed by the tensor product of matter and field components is given by 
\begin{eqnarray}\label{test}
|\alpha;\, h_1, \, \vec{\gamma}\rangle &=& \frac{{\rm e}^{-|\alpha|^2/2}}{\{ h_1; \, \vec{\gamma} \, | \, h_1; \, \vec{\gamma}\} ^{1/2}} \ \sum_{\nu=0}^\infty \sum_{n=0}^{h_1} \sum_{m=0}^{h_1-n}\frac{\sqrt{h_1!}\ \alpha^\nu \gamma_3^{n}\ \gamma_2^{m}  }{\sqrt{\nu!\  n!\ m!\ (h_1-n-m)!}}  \nonumber \\[3mm]
&&\times |\nu\, , h_1-n-m\, , n\, , m\rangle_F,
\end{eqnarray}
where, by means of (\ref{eq.Kp}) with $h_2=h_3=0$ and $\vec \gamma^\prime = \vec \gamma$, 
\begin{equation*}
\{ h_1, \, \vec{\gamma} \,  | \, h_1, \, \vec{\gamma}\}  = \left(1+|\gamma_2|^2+|\gamma_3|^2\right)^{h_1}
\end{equation*}
and one can thus write (\ref{test}) in the form
\begin{eqnarray}\label{eq.sc.coh}
|\alpha;\, h_1, \, \vec{\gamma}\rangle &=& \frac{{\rm e}^{-|\alpha|^2/2}}{\left(1+|\gamma_2|^2+|\gamma_3|^2\right)^{h_1/2}}\frac{1}{\sqrt{h1!}}\ {\rm e}^{\alpha\bm{a}^\dag}\left(\bm{b}_1^\dag + \gamma_3 \, \bm{b}_2^\dag + \gamma_2 \, \bm{b}_3^\dag\right)^{h_1} \nonumber \\[3mm] &&\times |0,0, 0,0\rangle_F \, .
\end{eqnarray}

To have a variational state with a definite total number of excitations, we replace the eigenvalue of the number of photons by $\nu =M- \lambda_2 n - \lambda_3 m$. To select the atom configuration one uses the corresponding values  of $\lambda_2$ and $\lambda_3$ in Table \ref{t1}. Then the unnormalized projected state is
\begin{eqnarray}\label{eq.proj}
|\alpha;\, h_1, \, \vec{\gamma}\}_M &=&  \sum_{n=0}^{h_1} \sum_{m=0}^{h_1-n}\frac{\sqrt{h_1!}\ \alpha^{M- \lambda_2 n - \lambda_3 m} \gamma_3^{n}\ \gamma_2^{m}  }{\sqrt{(M- \lambda_2 n - \lambda_3 m)!\  n!\ m!\ (h_1-n-m)!}}  \nonumber \\[3mm] &&\times |M- \lambda_2 n - \lambda_3 m, \ h_1-n-m,\ n,\ m\rangle_F,
\end{eqnarray}
%\end{widetext}
and contains only states with a fixed value of $M$, so that the semi-classical coherent state is written in simple form as
\begin{eqnarray}\label{eq.exp.proj}
|\alpha;\, h_1, \, \vec{\gamma}\rangle = \frac{{\rm e}^{-|\alpha|^2/2}}{\left(1+|\gamma_2|^2+|\gamma_3|^2\right)^{h_1/2}} \, \sum_{M=0}^\infty |\alpha;\, h_1, \, \vec{\gamma}\}_M \, . \quad 
\end{eqnarray} 
The state $|h_1;\alpha\vec{\gamma}\}_M$ is the unnormalized projected state.
%%%%%%

Since, the expectation value of the total number of excitations is very close to the exact one [see Fig. \ref{f10}(a)], one may correct the semi-classical ground state by considering, for each value of $\langle \bm{M}\rangle$, the corresponding projected state $|h_1;\alpha\ \vec{\gamma}\}_M$, but as the semi-classical calculation of $\langle \bm{M}\rangle$ is a continuous function of the control parameters, it is necessary to discretize it. We do this by defining $M_{dis} = \lceil\langle \bm{M}\rangle\rceil$, the ceiling of the expected $\bm{M}$ value. So, for particular values of the control parameters we define the {\it projected state} as $|h_1;\alpha\ \vec{\gamma}\}_{M_{dis}}$. 

We will use these projected states to calculate the expectation values of observables. To this end, the overlap is given by [from Eq. (\ref{eq.proj})]
%
%\begin{widetext}
\begin{eqnarray}\label{eq.norma.proj}
\{\alpha;\, h_1, \, \vec{\gamma} | \alpha;\, h_1, \, \vec{\gamma}\}_{M_{dis}} &=&  \sum_{n=0}^{h_1} \sum_{m=0}^{h_1-n}\frac{h_1!\ \rho_c^{2(M_{dis}- \lambda_2 n - \lambda_3 m)}   }{(M_{dis}- \lambda_2 n - \lambda_3 m)!} \nonumber \\[2mm]
&&\times \frac{ \varrho_{3c}^{2n}\ \varrho_{2c}^{2 m}  }{  n!\, m!\, (h_1-n-m)!}\,,
\end{eqnarray}
where we have evaluated at the critical points of the semi-classical calculation. As an example,  the unnormalized expectation value of the number of photons reads
\begin{eqnarray}\label{eq.nu.proj}
\{\alpha;\, h_1, \, \vec{\gamma} |\bm{n}| \alpha;\, h_1, \, \vec{\gamma}\}_{M_{dis}} &=&  \sum_{n=0}^{h_1} \sum_{m=0}^{h_1-n} \frac{h_1!\ \rho_c^{2(M_{dis}- \lambda_2 n - \lambda_3 m)}   }{(M_{dis}- \lambda_2 n - \lambda_3 m-1)!}
\nonumber \\
&&\times \frac{\varrho_{3c}^{2n}\ \varrho_{2c}^{2 m}  }{  n!\, m!\, (h_1-n-m)!}\, .
\end{eqnarray}
%\end{widetext}

Fig.~\ref{f11} shows, in similar form to Fig.~\ref{f10} and for the same parameters and atomic configuration, the expectation values of $\bm{M}$ [Fig.~\ref{f11}(a)],  $\bm{n}$ [Fig.~\ref{f11}(b)] and its fluctuations $\left(\Delta n\right)^2$ [Fig.~\ref{f11}(c)], comparing the exact calculation (EQ, darker dots) with the corresponding one using the projected state (SC, lighter dots). Notice that now the photon fluctuations provided by the projected state are comparable with the exact calculation, showing that the projected state corrects the wrong behavior of the fluctuations of the standard coherent state.

Figure~\ref{f11} is shown for $N_a=5$ atoms; for larger values of $N_a$ both calculations will be indistinguishable.

% Figura 13
\begin{figure*}
\begin{center}
\includegraphics[width=0.48\linewidth]{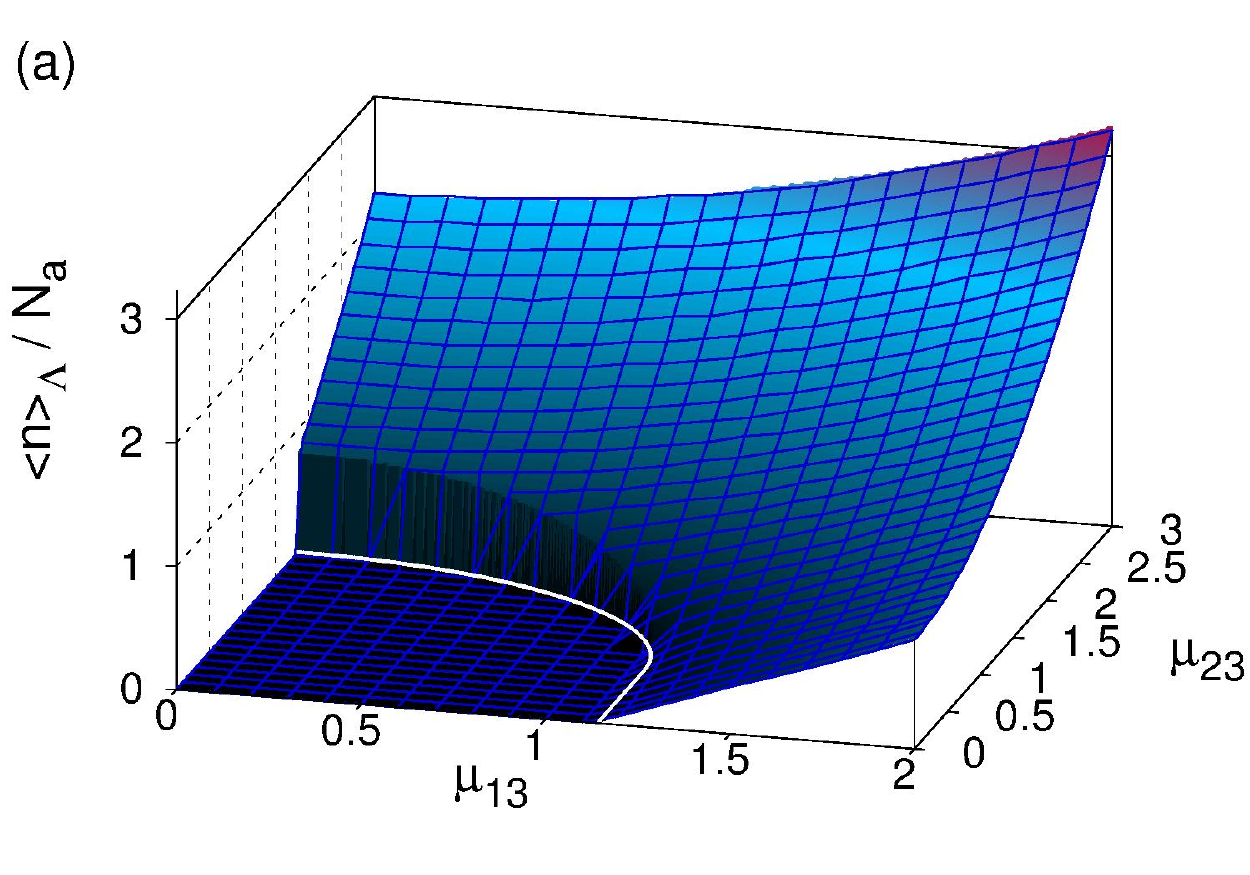} \,
\includegraphics[width=0.48\linewidth]{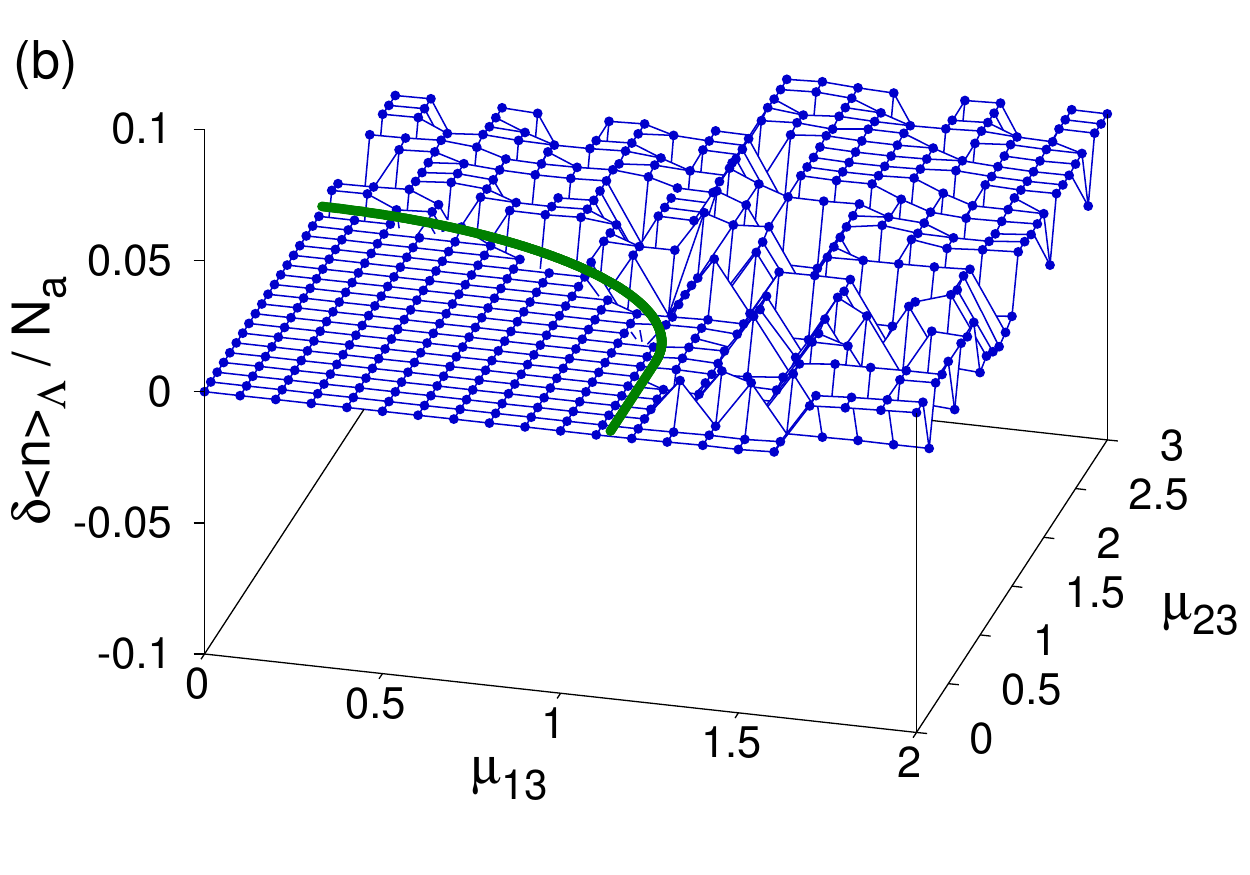}
\end{center}
\caption{(Color online.) Exact and projected solutions compared for atoms in the $\Lambda$ configuration in a non-resonant condition $\Delta_{31}=0.3,\ \Delta_{32}=-0.2$, considering $N_a=40$ atoms. (a) expectation value of the number of photons, showing no visual differences, and (b)  the corresponding fluctuations. In both cases, the quantities were normalized to the number of atoms $N_a$. }\label{f13}
\end{figure*}

%

% Figura 14
\begin{figure*}
\begin{center}
\includegraphics[width=0.48\linewidth]{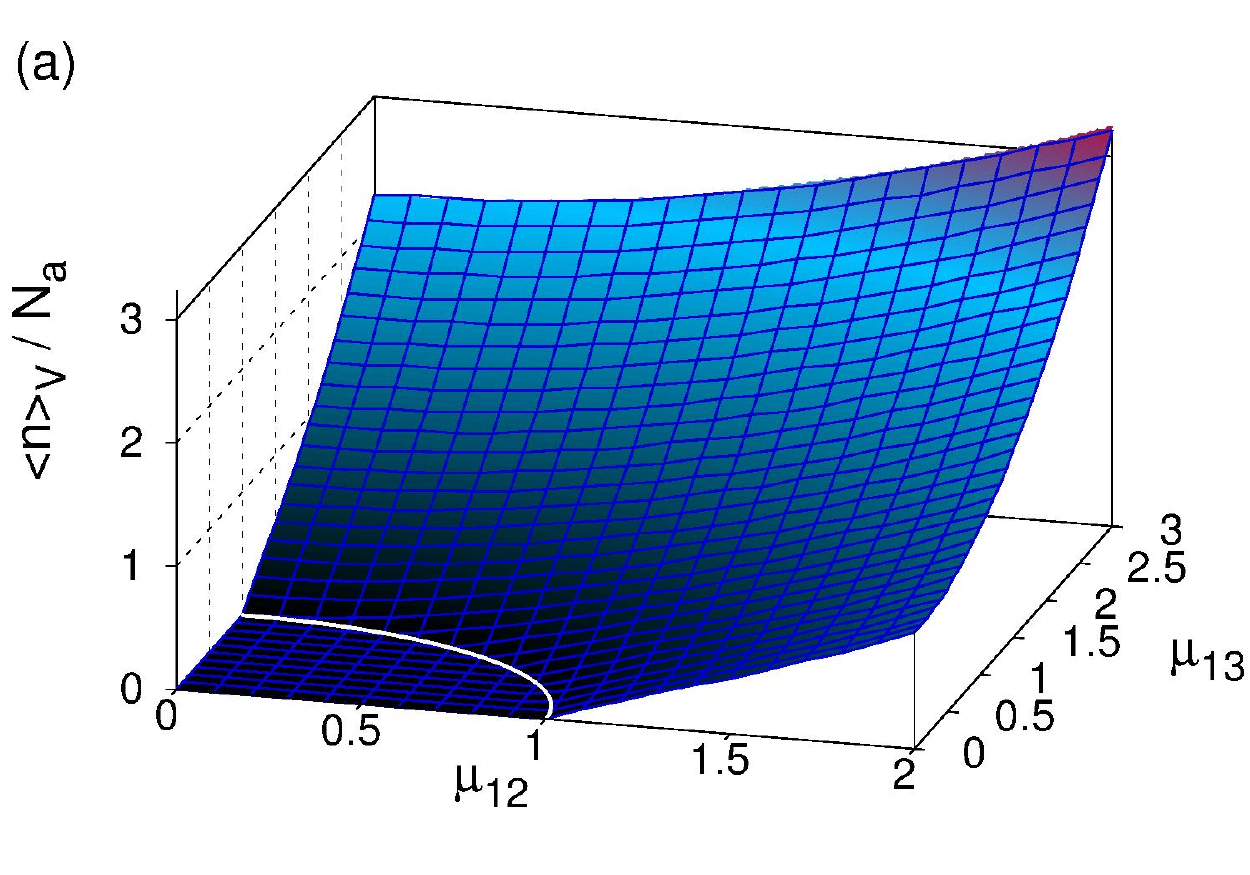} \,
\includegraphics[width=0.48\linewidth]{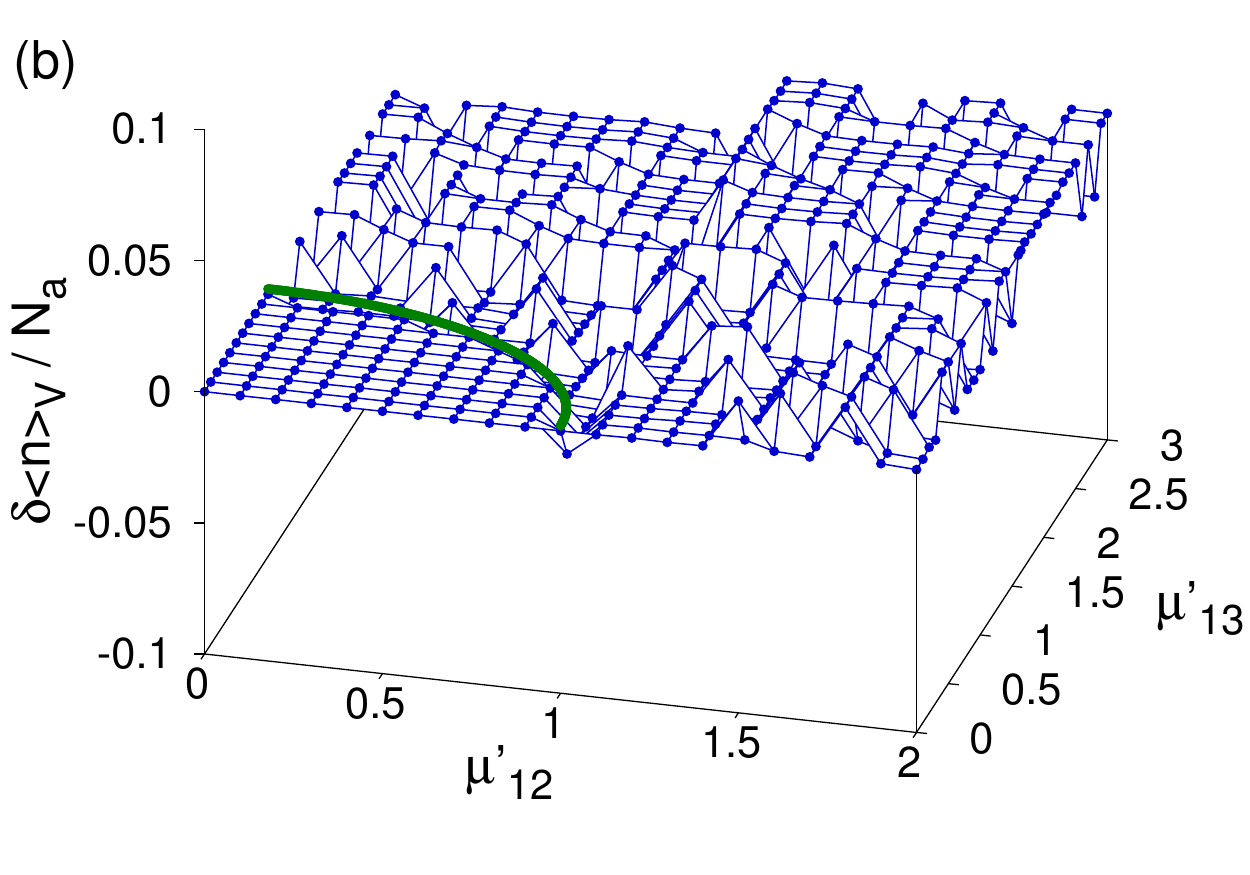} 
\end{center}
\caption{
(Color online.) Exact and projected solutions compared for atoms in the $V$ configuration in double resonance $\Delta_{21}=\Delta_{31}=0$, considering $N_a=40$ atoms. (a) expectation value of the number of photons, showing no visual differences, and (b) the corresponding fluctuations. In both cases, the quantities were normalized to the number of atoms $N_a$. }\label{f14}
\end{figure*}

\subsection{$\Xi$ configuration}\label{quantum.proj.X}

For $N_a=40$ atoms in the $\Xi$ configuration, in double resonance, i.e., $\Delta_{21}=\Delta_{32}=0$, the expectation value of the number of photons and its fluctuations are compared for both the exact (mesh) and projected variational (continuous surface) states in Fig.~\ref{f12}. For $\langle\bm{n}\rangle_\Xi /N_a$ [Fig.~\ref{f12}(a)] there are no visual differences. In fact, this figure is identical to figure~\ref{f3} where the expectation value of the number of photons is shown for the semi-classical coherent state. Table \ref{t2} shows the minimum and maximum values of the difference between the projected and exact results given by

\begin{table}%[h]
\caption{Maximum and minimum difference between projected and exact quantum results Eq. (\ref{eq.deltaNph}) for the three configurations of the atom. The maximum difference is reached close to the separatrix due to the finite number of atoms; this value diminishes as we move away from the separatrix or as $N_{a}$ is increased.}
\begin{center}
\begin{tabular}{|c|cc|}
\hline
Configuration&$\min\delta \langle{\bf n}\rangle/N_a$& $\max\delta \langle{\bf n}\rangle/N_a$\\
\hline  & &\\[-3mm]
$\Xi$&0&$\sim3.3\times10^{-2}$\\
$\Lambda$&0&$\sim7.7\times10^{-1}$\\
$V$&0&$\sim2.4\times10^{-2}$\\
\hline
\end{tabular} 
\end{center}
\label{t2}
\end{table}

\begin{eqnarray}\label{eq.deltaNph}
\frac{\delta\langle  \bm{n}\rangle}{N_a} &\equiv& \left|\frac{\langle \bm{n}\rangle_{proj}-\langle \bm{n}\rangle_q}{N_a}\right|,
\end{eqnarray}
in absolute value and normalized by the number of atoms. In the normal regime the difference vanishes exactly, while in the collective regime it is of order $\sim 10^{-2}$. Finally, Fig.~\ref{f12}(b) shows the corresponding fluctuations presenting very small differences in the collective regime.

\subsection{$\Lambda$ configuration}\label{quantum.proj.L}

For the $\Lambda$ configuration we consider a non-resonant case $\Delta_{31}=0.3$ and $\Delta_{32}=-0.2$, and $N_a=40$ atoms. Under these conditions the behavior of the physical observables resembles that of the $\Xi$ configuration by showing  both, first- and second-order phase transitions.

Fig.~\ref{f13}(a) shows the comparison between the expectation values of the number of photons calculated with respect to the exact (mesh) and projected (continuous surface) states, where one may observe an excellent agreement between both surfaces.  In the normal regime the difference $\delta\langle  \bm{n}\rangle/N_a $ vanishes exactly, while in the collective regime the maximum value is of order $\sim 10^{-1}$. As in the previous case, this diminishes as we move away from the separatrix or as $N_{a}$ is increased.  Fig.~\ref{f13}(b) compares the fluctuations in the number of photons. In contrast to the $\Xi$ configuration, here the fluctuations tend asymptotically to a constant value.

\subsection{$V$ configuration}\label{quantum.proj.V}

Finally, we consider the expectation value of the number of photons for atoms in the $V$ configuration, in a double resonance condition $\Delta_{21}=\Delta_{31}=0$, with $N_a=40$ atoms. As discussed in the semi-classical calculation of Sec.~\ref{NR.V}, the qualitative behavior of the physical quantities for this configuration is independent of the detuning considered. 

Fig.~\ref{f14}(a) shows the comparison between the expectation value of the number of photons evaluated for the exact quantum (mesh) and projected (continuous surface) states. One may observe that there are no visual differences. Differences of order $\sim 10^{-2}$ appear in the collective regime,    as shown in Table.~\ref{t2}. The fluctuations are shown in Fig.~\ref{f14}(b), and once again these approach a constant in the collective regime, in a similar fashion to the $\Lambda$ configuration.

\section{Concluding remarks}\label{concluding}

The ground state of a system of $N_a$  three-level atoms interacting via dipole interactions with a one-mode quantized electromagnetic field was described, in the rotating wave approximation. The different atomic configurations $\Xi$,
$\Lambda$, and $V$ were considered. 

The ground state was approximated by a test function (semi-classical state) constructed from the tensorial product of Heisenberg-Weyl and U(3) coherent states. There are two different behaviors called normal, where the ground state is given by all the atoms in the lower energy level and without photons (${\cal M}=0$), and collective, where the atoms are distributed amongst the three levels of the system, and with a corresponding number of excitations ${\cal M} \neq 0$ and average number of photons $\langle \bm n \rangle \neq 0$.

The ground state of the system in the $\Xi$ configuration exhibits first- and second-order transitions, independently of the detuning values (see Fig.~\ref{f1}). For atoms in the $\Lambda$ configuration, one finds for equal detuning values that it can only present second-order transitions, this is shown analytically in Eq.~(\ref{eq.L.minE}). For different detuning parameters, this configuration yields first- and second-order transitions (see Fig.~\ref{f4}). For atoms in the $V$ configuration, independently of the detuning, there are only second-order transitions, and this is shown analytically in Eq.~(\ref{eq.V.minE}) for equal detuning parameters and numerically in Fig.~\ref{f7} for other cases. 

For all atomic configurations, we have found that in the normal regime the expectation value of the total number of excitations with respect to the ground state is zero and it follows a Poissonian distribution. In the collective regime the total number of excitations for the ground state has a sub-Poissonian distribution as shown in Figs.~\ref{f2}, \ref{f5} and \ref{f8}. The expectation values of the number of photons given in Figs.~\ref{f3} and \ref{f6} display discontinuities where first-order transitions take place.

For $N_a=5$ atoms in the $\Xi$ configuration, the exact quantum calculation for the expectation values of the total number of excitations and of the number of photons were compared with the corresponding semi-classical ones. Both calculations agree, as shown in Figs.~\ref{f10}(a) and (b). Similar results can be obtained for the other configurations. However, the fluctuations in the number of photons are very different [Fig.~\ref{f10}(c)], which suggests to consider a new test function. We proposed to project the semi-classical test function to a definite total number of excitations ${\cal M}$; this projected state was obtained by choosing the ceiling value of ${\cal M}^c$ together with the critical points for the semi-classical case. We showed that the photon fluctuations provided by the projected state are comparable with those of the exact calculation, so that the projected state corrects the wrong behavior of the standard coherent state (cf. Fig.~\ref{f11}). 

Finally, for $N_a=40$ atoms the expectation values and fluctuations of the number of photons were calculated. In all cases, we have found that the results for the projected state are indistinguishable from those of the exact one as can be seen in Figs.~\ref{f12}(a), \ref{f13}(a) and \ref{f14}(a). To have a quantitative estimation of the differences between these calculations, we used Eq.~(\ref{eq.deltaNph}) observing the major differences along the separatrix [Table~\ref{t2}]. This is valid for all atomic configurations. 

For the $\Xi$ configuration, in the double resonance case and any number of atoms, we found a fixed point in the parameter space $(\mu_{12}=1, \, \mu_{23}=\sqrt 2)$, in which there is coexistence between three different eigenstates associated to the same energy. They correspond to a total number of excitations of ${\cal M}=0$, ${\cal M}=1$, and ${\cal M}=2$, thus implying the presence of a triple point in the parameter space. 

When more than one electromagnetic modes are present the physics can be much richer. Specific cases where each mode resonates with one and only one atomic energy transition, and where one considers only one atomic configuration, have been studied in the thermodynamic limit~\cite{brandes}. The general situation, however, is highly non-trivial and merits further study.

\subsection*{Acknowledgments}
This work was partially supported by CONACyT-M\'exico (under project
101541), and DGAPA-UNAM (under project IN102811).

\appendix

\section{Matrix elements of the ${\rm U}(3)$ operators}\label{ap.Aij}

The matrix elements of the generators of ${\rm U}(3)$, for a general irreducible representation $[h_1,h_2,h_3]$, can be found in~\cite{moshinsky67}. For the totally symmetric representation, $[h_1,0,0]$, the Gelfand-Tsetlin states take the form
\begin{eqnarray}
|qr\rangle&\equiv&\left|\begin{array}{c c c} q & & 0 \\ & r & \end{array} \right\rangle,
\end{eqnarray} 
where $q$ and $r$ take values from $0$ to $h_1$. 

In this representation, the matrix elements of the atomic operators $\bm{A}_{ij}$ are given by
%
%\begin{subequations}
\begin{eqnarray}
\langle  q r|\bm{A}_{11}|q r\rangle = r,
\end{eqnarray}
\begin{eqnarray}
\langle  q r|\bm{A}_{22}|q r\rangle = q-r,
\end{eqnarray}
\begin{eqnarray}
\langle  q r|\bm{A}_{33}|q r\rangle = h_1-q,
\end{eqnarray}
\begin{eqnarray}
\langle  q r+1|\bm{A}_{12}|q r\rangle = \sqrt{(q-r)(r+1)}, \qquad
\end{eqnarray}
\begin{eqnarray}
\langle  q+1 r+1|\bm{A}_{13}|q r\rangle = \sqrt{(h_1-q)(r+1)}, \qquad
\end{eqnarray}
\begin{eqnarray}
\langle  q+1 r|\bm{A}_{23}|q r\rangle = \sqrt{(h_1-q)(q-r+1)}, \qquad
\end{eqnarray}
%
%\end{subequations}
and zero for other cases.

%
%\bibliographystyle{vancouver}%{iopart-num}%
%\bibliography{icn}
%

\end{document}